\documentclass[preprint,12pt,authoryear]{elsarticle}
\usepackage[margin=0.7in]{geometry}
\usepackage{color}

\usepackage[utf8]{inputenc}
\usepackage{graphicx}
\usepackage{amssymb}
\usepackage{amsthm}
\usepackage{tabularx}
\usepackage{bm}
\usepackage{multirow}
\usepackage{lineno}
\usepackage{tabularx}
\usepackage{amsfonts}
\usepackage{amsmath}
\usepackage{array}
\usepackage{framed}
\usepackage{appendix}
\usepackage{todonotes}

\newtheorem{theorem}{Theorem}[section]
\newtheorem{proposition}[theorem]{Proposition}

\theoremstyle{definition}

\newtheorem{remark}[theorem]{Remark}
\newcommand{\teps}{\bm{\varepsilon}}

\newcommand{\R}{\mathbb{R}}

\usepackage{natbib}
\usepackage{graphicx}

\newcommand\EEE{\color{black}}

\newcommand{\mk}{\color{black}}

\journal{International Journal of Solids and Structures}

\begin{document}

\begin{frontmatter}

\title{Thermomechanical model for NiTi-based shape memory alloys covering macroscopic localization of martensitic transformation}

\author[UT,UJF]{M. Frost\corref{cor1}}
\ead{mfrost@it.cas.cz}
\author[UT,MUUK]{B. Bene\v{s}ov\'{a}}
\author[UT]{H. Seiner}
\author[UTIA]{M. Kru\v{z}\'{i}k}
\author[FzU,UJF]{P. \v{S}ittner}
\author[UT,UJF]{P. Sedl\'{a}k}

\cortext[cor1]{Corresponding author. Institute of Thermomechanics, Dolej\v{s}kova 5, CZ-18200 Prague. Tel.: +420 266 053 712. Fax: +420 286 584 695.}

\address[UT]{Czech Academy of Sciences, Institute of Thermomechanics, Dolej\v{s}kova 5, 18200 Prague, Czech Republic}
\address[UJF]{Czech Academy of Sciences, Nuclear Physics Institute, Husinec - \v{R}e\v{z} 130, 250 68, \v{R}e\v{z}, Czech Republic}
\address[MUUK]{Department of Mathematical Analysis, Charles University, Sokolovsk\'{a} 83, 18600 Prague, Czech Republic}
\address[UTIA]{Czech Academy of Sciences, Institute of Information Theory and Automation, Pod Vod\'{a}renskou v\v{e}\v{z}\'{i} 4, 18200, Prague, Czech Republic}
\address[FzU]{Czech Academy of Sciences, Institute of Physics, Na Slovance 2, 18221 Prague, Czech Republic}

\begin{abstract}
The work presents a thermomechanical model for polycrystalline NiTi-based shape memory alloys developed within the framework of generalized standard solids, which is able to cover loading-mode dependent localization of the martensitic transformation. The key point is the introduction of a novel austenite-martensite interaction term responsible for strain-softening of the material. Mathematical properties of the model are analyzed and a suitable regularization and a time-discrete approximation for numerical implementation to the finite-element method are proposed. Model performance is illustrated on two numerical simulations: tension of a superelastic NiTi ribbon and bending of a superelastic NiTi tube.

\end{abstract}




\end{frontmatter}

\section{Introduction}

Having found many applications in medicine, civil engineering or aerospace industry \citep{JAN-GIB}, NiTi-based alloys are a prominent class within shape memory alloys (SMAs) usually utilized in the form of thin structures, e.g. wires, strips, tubes or plates. It is often experimentally observed that stress-induced martensitic transformation (mainly when it is induced by uniaxial tension) does not occur in a spatially homogeneous manner on \mk  a \EEE macroscopic scale; instead, localized “martensitic bands" appear within the austenitic sample and the transformation propagates by their multiplication and/or by movement of their fronts. The macroscopic picture is very similar to the localization of plastic deformation in certain steels and alloys well-known as Lüders bands: the onset of transformation is usually accompanied by a stress overpeak followed by a stress plateau, and \mk the \EEE material rehardens after exhausting the available portion of transformation strain.

The simplest example of localization can be found in NiTi wires loaded in tension \citep{SHA-KYR, SCIENCE}. A wider range of localization patterns can be observed in tensile deformation on NiTi thin strips as thoroughly reported by \cite{SHA-KYR-97,SHA-KYR-97b}. Their works also documented the strong dependence of localization patterns on \mk the \EEE deformation rate. The dynamics of formation, propagation and coalescence of transformation bands on NiTi strips and the relation to the strain-rate was further studied \citep[e.g. by][and many others]{PIE-TOB-13,ZHA-SUN-10,BIA-PER-18}.

A further sample shape that can be easily manufactured and exhibits various localization patterns is the tube. \citet{SUN-LI-02} were first to show spiral bands forming in tension on the surface of thin-walled superelastic NiTi tubes. The development of localized deformation in tension and homogeneity of deformation in compression were later documented by \citep{MAO-LIU, REE-DAL-14, BEC-KYR-14}. Bent tubes exhibit wedge-like localization patterns on the part of the surface deformed predominantly in tension whereas homogeneous deformation was detected on the part deformed predominantly in compression \citep{REE-DAL-14,BEC-KYR-14}. Experiments were also performed in multiaxial loading: propagating transformation fronts were observed in pure tension, they were absent in pure torsion, and progressive behavior in between these pure modes were observed in \citep{SUN-LI-02,REE-DAL-17}; in proportional biaxial loading experiments by \citet{BEC-KYR-16}, localized helical bands with inclinations dependent on the stress ratio formed except for the loading modes close to equibiaxial tension. More detailed review on experimental results on the localization and their relation to constitutive modeling challenges can be found in \citep{APPA-LOC}.

It is expected that the key for understanding of formation of macroscopic martensitic bands are interactions at the mesoscopic level, i.e. the level of transforming grains (where the martensitic transformation is always localized) and their aggregates \citep{SIT-LIU}. This motivated several in-situ microstructure studies focused on characterization of stress redistributions associated with martensite nucleation and growth \citep{YOU-EGG,SCIENCE}. A detailed picture of the martensitic band front in a loaded NiTi wire obtained by 3D X-ray synchrotron diffraction \citep{SCIENCE} shows that the internal stress states in grains massively change and redistribute within the propagating front so that, at the onset of transformation, austenitic grains were (in average) exposed to equivalent stresses more than 200 MPa higher then externally measured plateau stress. The observed stress heterogeneity was rationalized with help of a FEM simulation assuming considerable strain-softening during progress of martensitic transformation. Since strain-softening is “invisible" in macroscopic stress-strain curves with a plateau, dedicated experiments revealing the course of the strain-softening are needed. Quantitative results have been reached by \citet{HAL-KYR} via a sophisticated measurement of a NiTi-stainless steel composite and, recently, by \citet{ALA-SIT-17} via a special geometry of NiTi bulk specimen.

Despite abundance of constitutive models of SMAs in the literature -- originating at different scales of description, see recent reviews by \citet{ZAK-BZ-16,ZAK-BZ-16b} -- comprehensive three-dimensional macroscopic (continuum) models incorporating localization have been attempted rarely. Most frequently, simple isotropic plasticity-based models with the strain-softening are used to study localization in NiTi wires \citep{IAD-SHA,BAD-KAD}, strips \citep{SHA-KYR-97b,AZA-RAJ-07} or tubes deformed in tension \citep{JIA-KYR-LAN-17b,MRH-STU-19}.

Involving the strain-softening into constitutive models brings inevitably problems with their regularity such as mesh-dependence of solution or spurious concentration of strain to infinitely small regions. These problems were already analyzed (motivated by their practical impact mainly in simulation of damage) and several regularization techniques were proposed \citep[see e.g.][]{BAZ-JIR-02, JIR-ROL-03}. They are based on so-called nonlocal continuum theories, in which response of a material point is not uniquely determined by values of state and internal variables (fields) in that point only, but the state of material points in its vicinity is also taken into account. 
A useful tool for incorporating such information is introduction of so-called non-local variable(s). Then, two issues have to be addressed: (a) defining how the nonlocal (“twin”) variable is related to a local one(s), and (b) modifying the constitutive laws, which involve both local and non-local variables. Concerning the first point, two well-established approaches can be borrowed from structural mechanics: the implicit nonlocal gradient approach (iNGA) and the nonlocal integral approach (NIA). In iNGA, the local and corresponding nonlocal variables are linked via an additional (elliptic) partial differential equation; in NIA, these variables are linked via an integral relation -- the nonlocal variable in a material point is defined as a weighted integral average of the values of the local one gained in a close neighborhood of that point. Corresponding general mathematical formulations are closely related, since iNGA can be derived from NIA using particular weighting functions \citep{PEE-GEE-01}. Both approaches naturally incorporate an internal parameter related to some characteristic length scale related to the extend of relevant neighborhood. The examples of iNGA in SMA models can be found in \citep{DUV-CHE, ARM-BZ-14, BAD-KAD}, whereas NIA approach was employed e.g. in \citep{AHM-HOS, SCIENCE}.

Recently, a few more elaborated models which tried to capture also the loading mode-dependence of localization patterns observed in experiments appeared \citep{POU-WAG-17,JIA-KYR-LAN-17,JIA-KYR-LAN-17b}. Namely, they attempt to capture the fact that whereas the transformation in tension is usually localized, the transformation in compression is observed to be homogeneous \citep{ELI-WAG,WAT-REE-18} via heuristic modifications of the ``yield criterion'' used in the plasticity-based models.

In this work, we present an extension of our original SMA model formulated in the framework of generalized standard solids by \citet{SED-FRO-IJOP}. We introduce a novel austenite-martensite interaction term and show the capability of the extended model to reproduce evolution of localized martensitic transformation in NiTi SMAs. The original model \citep{SED-FRO-IJOP} is briefly summarized in Section \ref{sec-local}. The derivation of the interaction term, which is based on the elastic energy of a material with misfitting inclusions by \citet{MOR-TAN-73}, is presented in Section \ref{sec-nonloc}. The derivation of the interaction term is not limited to any particular loading mode emphasizing the ambition of the model to be used in general loading scenarios. Splitting the internal variables into local and non-local ones, which is done heuristically in the definition of the interaction energy, allows to rigorously perform the regularization of the complete model as presented in Section \ref{sec-maths}. Numerical implementation into finite element method (FEM) is described and illustrating simulations are performed in Section \ref{sec-num}.

\section{Local model}
\label{sec-local}

Within this work, we develop a phenomenological constitutive model of the NiTi SMAs capable to capture the localization effects. The term phenomenological here means that the model aims to describe the polycrystalline material in some average sense and underlying microscopic features are taken into account through \emph{inner variables} of the model. Such models allows for an easy implementation, less time consuming simulations and have the potential to be applied in industrial applications. Thus, a large number of such models has been proposed so far \citep[e.g.][and many others]{LAG-CHE,STU-PET-12,STE-BRI,AUR-BON-14,MEH-ELA-14b,GU-ZAK,CHA-CHE-16}.

Here, we adapt a three-dimensional macroscopic model for NiTi SMA proposed in \citep{SED-FRO-IJOP} and further developed and validated in \citep{JMEP-STENT,FRO-BEN-MMS,JIMSS-SPRING,SCIENCE,SMS-SNAKE} that has been shown to perform well even for non-proportional loading. We review the model here for the reader's convenience, as we will extend it later in Section \ref{sec-nonloc} to capture localization.

The modelled specimen is assumed to occupy, at the reference configuration, the domain $\Omega \subset \R^3$. As the primary state variable we choose the displacement of the specimen $u:\Omega \to \R^3$. The total strain tensor, $\teps(x)$, is related to the displacement via
$$
\teps(x) = \frac{1}{2} \big(\nabla u(x) + (\nabla u(x))^\top \big),
$$
which is given locally in every material point $x \in \Omega$. As frequent in macroscopic modelling of SMAs \citep[cf.][]{GU-ZAK, SAD-BHA-3D, CHE-DUV}, we additionally introduce two internal variables: the scalar $\xi: \Omega \to \R^3$ representing the volume fraction of martensite and satisfying $\xi(x) \in [0,1]$ in every material point and a tensor variable, $\teps^{\rm tr}: \Omega \to \R^3$, representing the transformation strain. With these two variables, we may write the the conventional small strain decomposition, assumed to be valid for every $x \in \Omega$, as
\begin{equation}
\mk \teps^{\rm el}(x)\EEE  = \teps(x) - \xi(x) \teps^{\rm tr}(x).
\end{equation}
\mk Here $\teps^{\rm el}$ is  the elastic strain and  $\teps^{\rm tr}$ is a macroscopic variable \EEE
used for storing information about microscopic internal structure of martensite. Later, particularly in the mathematical part, we will also use the inelastic strain defined via
\begin{equation}
\teps^{\rm in}(x) = \xi(x) \teps^{\rm tr}(x).
\label{def-inelastic-strain}
\end{equation}

Crystallographic considerations show \citep{OTS-WAY} that there exists a maximum value of strain that is attainable due to phase transformation; so, the transformation strain is considered to lie in a bounded convex set. In addition, the austenite-martensite transition is nearly volume preserving, so that it is justified to consider $\teps^{\rm tr}$ (as well as $\teps^{\rm in}$) to be trace-free tensors. For the model in hand, this means that for every material point $x \in \Omega$, we require
\begin{equation}\label{eq:eps-tr-def}
\teps^{\rm tr}(x) \in \left\{ A \in
\mathbb{R}^{3\times 3}:  A \text{ is symmetric, } {\rm
tr}( A) = 0,\; \langle  A \rangle \leq 1 \right\},
\end{equation}
where ${\rm tr}( A)$ denotes the trace of a tensor $ A$
and $\langle \cdot \rangle: \mathbb{R}^{3 \times 3} \to
\mathbb{R}^+$ is a suitable positively 1-homogeneous convex
function; by a particular form of this function,
tension-compression asymmetry is captured in the model.

It is convenient to formulate the model within the framework of so-called \emph{generalized standard solids} \citep[see][]{HAL-NGU}. That means that we need to prescribe two scalar functions, the free energy $f_T(\teps, \teps^{\rm tr}, \xi)$, that depends on the state variables, as well as dissipation function $d_T( \teps^{\rm tr}, \xi, \dot{\teps}^{\rm tr}, \dot{\xi})$, that depends both on the internal state variables and their rates. The free energy and the dissipation function may depend on the temperature $T$ that we, however, consider prescribed in the whole specimen (quasistatic approximation), which is indicated by the respective index.

Within this section, the free energy and dissipation function are understood to be given \emph{locally} in each material point (so that all the variables actually depend on $x \in \Omega$), but for a better readability we suppress to indicate this from now on.

The free energy is given as a sum of elastic and chemical contributions:
\begin{align}
f^{T}(\varepsilon,\xi,\varepsilon^\mathrm{tr})&=\underbrace{\frac{1}{2}(\varepsilon-\xi \varepsilon^\mathrm{tr}):\mathbb{C}(\xi):(\varepsilon-\xi \varepsilon^\mathrm{tr})}_\text{elastic energy}+\underbrace{\Delta s^{AM}(T-T_{0})\xi}_\text{chemical energy},
\label{eq-energy}
\end{align}
where $\mathbb{C}$ is the tensor of elastic constants, $s^{AM}$ is the specific entropy difference between the austenite and martensite phase and $T_0$ is the temperature at which austenite and martensite are (energetically) at equilibrium.

Furthermore, we choose the dissipation function as
\begin{align*}
 d^T(\varepsilon^{\rm tr},\xi, \dot{\varepsilon}^{\rm tr},\dot{\xi})
  =  \begin{cases}
  \Delta s^{AM}\left[T_0 - M_{\rm s} + \xi (M_{\rm s} - M_{\rm f})\right]|\dot{\xi}|
  +\; \sigma^{\rm reo}(T)\|\xi \dot{\varepsilon}^\mathrm{tr}+\dot{\xi}\varepsilon^\mathrm{tr} \| & {\rm if }\; \dot{\xi} \geq 0, \\
  \Delta s^{AM}\left[A_{\rm f} - T_0 + \xi (A_{\rm s} - A_{\rm f})\right]|\dot{\xi}|
  +\; \sigma^{\rm reo}(T)\left[\|\dot{\xi}\varepsilon^{\rm tr}\| +
  \|\xi \dot{\varepsilon}^\mathrm{tr}\|\right] & {\rm if}\; \dot{\xi} < 0,
  \end{cases}
\end{align*}
where $M_\mathrm{s}, M_\mathrm{f}$ as well as $A_\mathrm{s}, A_\mathrm{f}$ are the temperature at which the \emph{austenite-to-martensite (forward)} as well as the \emph{martensite-to-austenite (reverse)} start or finish, respectively. Moreover, $\sigma^{\rm reo}$ is a constant (may depend on temperature) that characterizes the amount of dissipated energy due to reorientation.

We refer the reader to \citep{SED-FRO-IJOP,FRO-BEN-MMS} for a detailed derivation of the form of this dissipation function. Let us just mention at this point that the form is chosen in such a way that reflects the following ideas: during the forward transformation, the appearing martensite can reorient immediately and thus the appearance of martensite and its reorientation are fully coupled processes. On the other hand, the  reverse transformation can only happen if the martensite can reorient to a form that can be coupled to austenite. Thus, a suitable reorientation of martensite must become favourable in order to allow the reverse phase transformation to occur.

\section{Non-local model}
\label{sec-nonloc}

In this section, we introduce an energy term which allows to capture localization into the constitutive model summarized in Section \ref{sec-local}. We use  micromechanics-inspired approach motivated by available (experimental) knowledge. The basic observations on localization of martensitic transformation in NiTi alloys can be summarized in the following points:

\begin{itemize}
\item Localization of martensitic transformation appears only in some loading modes. The localization was documented on NiTi wires, strips or tubes loaded
in tension, but most of the experiments on loading in compression and shear reveal macroscopically homogeneous transformation \citep{ELI-WAG,WAT-REE-18}.\footnote{The situation in torsion of NiTi tubes is ambiguous. Observations by \citet{SUN-LI-02} suggest homogeneity of transformation, whereas detailed SEM analysis in \citep{PEN-FAN-08} reveals mesoscopic heterogeneity in martensite distribution in the form of microscopic martensitic lamellae. Nevertheless, the heterogeneities are on a much finer scale compared to those occurring during tensile loading.}

\item The localization of martensitic transformation in tension cannot be explained as a sole result of geometry (cross-section) change associated with large deformation induced by transformation. 
Although the geometric changes, the decrease of the integral force due to necking in the transformed zone as well as the lack of material hardening\footnote{We recall here the well-known Consid{\`e}re criterion.} contribute to the observed asymmetry between the localized transformation in tension and the homogeneous one in compression,
several recent experiments \citep{HAL-KYR,SCIENCE,ALA-SIT-17} clearly documented a considerable stress decrease in the material during progression of martensitic transformation in tension. The decrease of stress depends on the particular material microstructure, and in magnitude, it could be even comparable with the transformation plateau stress.

\item The transformation stress in tension does not decrease linearly with the extent of transformation (martensite volume fraction). Experiments suggest the decrease is maximal at the beginning of the transformation, and then it continuously attenuates during further transformation \citep{HAL-KYR,ALA-SIT-17}.

\item On the level of single grains forming the polycrystalline material, the martensitic transformation spreads heterogeneously. Bands of martensite running through austenitic grains are observed during stress-induced transformation \citep{OTS-WAY}. The reason for the strain-softening can be thus sought in the interaction of fully martensitic regions with the surrounding elastic austenite.

\item On the other hand, temperature induced transformation in polycrystalline samples does not occur at a single transformation temperature
but in a broader temperature interval. This suggests there is a distribution of transformation temperatures within the polycrystalline material, and, consequently, also a distribution of transformation stresses. Moreover, differently oriented grains can fulfill transformation conditions at different stresses due to orientation dependence of transformation strain \citep{SIT-NOV}. Thus, we can expect that the forward transformation in polycrystalline samples induced by stress will start in the most favorably oriented grains (with the highest transformation temperature) and it will advance to other parts only with increasing loading.
\end{itemize}

In the next paragraph, we will explain the loading mode-dependence of localization of martensitic transformation in NiTi as the result of an interplay of hardening-like effects given by finite distributions of transformation strains and temperatures in a polycrystalline material and softening effects coming from the heterogeneous formation of martensite on the single-grain level and interaction of highly deformed martensitic zone with surrounding elastic austenite. The hardening during martensitic transformation was already incorporated in the original model by \citet{SED-FRO-IJOP} via the assumption of finite intervals of both forward and reverse transformations defined by transformation temperatures $M_{\rm s},M_{\rm f},A_{\rm s}$ and $A_{\rm f}$; the austenite-martensite interaction term is derived in the next section.

$\vphantom{}$

\noindent {\bf Energy of austenite-martensite interaction}

$\vphantom{}$

\noindent Derivation of austenite-martensite interaction term on macroscopic, phenomenological level is usually based on the Eshelby's solution of elastic field of ellipsoidal inclusion in elastic matrix \citep{ESH-57} and the derivation of average stress in material with misfitting inclusions by \citet{MOR-TAN-73}.

The elastic energy per unit volume of the specimen containing elliptical
inclusions with transformation strain $\teps^{\rm tr}$
reads as \citep{MOR-TAN-73}:

\begin{equation}
E^{\rm int}=-\frac{1}{2}\xi(1-\xi)\boldsymbol{\sigma}^{I\infty}:\teps^{\rm tr},
\end{equation}
where $\xi$ is the volume fraction of inclusions and $\boldsymbol{\sigma}^{I\infty}$ is
stress within a single inclusion existing in an infinitely extended body. Considering simple spherical inclusions, identical isotropic
elasticity of both austenite and martensite and the constraint $\mathrm{tr}(\teps^{\rm tr})=0$, the energy term condensates into a simple form:

\begin{figure}
\includegraphics[scale=0.6]{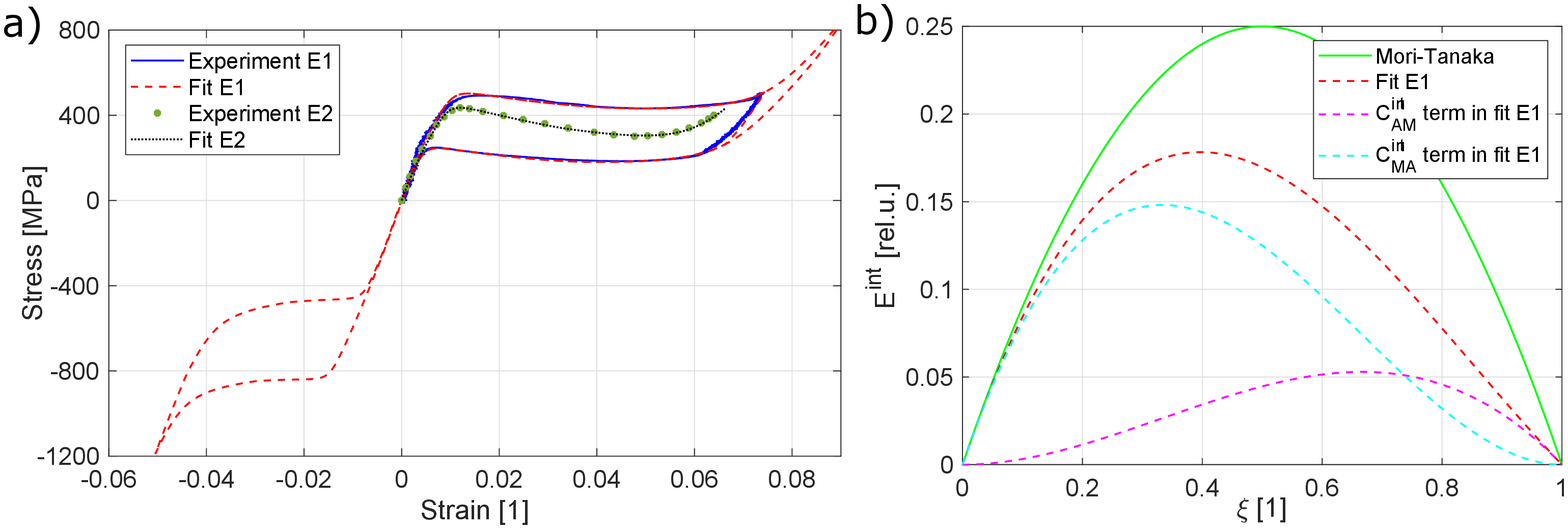}
\caption{\textbf{(a)} Experimental data on strain-softening in NiTi alloys by \citet{ALA-SIT-17} -- denoted E1 -- and \citet{HAL-KYR} -- denoted E2 -- plotted together with their best numerical fit reached by the non-local constitutive model introduced in the Section \ref{sec-nonloc}. \textbf{(b)} Character of the evolution of the internal elastic energy $E^{\rm int}$ with variation of the volume fraction of martensite $\xi$. For $C_{\rm MA}^{\rm int} = C_{\rm AM}^{\rm int}$ the  classical Mori-Tanaka symmetric case is recovered (marked in green). The symmetry of $E^{\rm int}$ is lost in the best fit of E1 (red dashed line) since $C_{\rm MA}^{\rm int} = 2.8\,C_{\rm AM}^{\rm int}$, see the respective contributions of corresponding energy terms (cyan and magenta dashed lines).}
\label{fig:uniaxial-fitting}
\end{figure}

\begin{equation}
E^{\rm int}=C^{\rm int}\xi(1-\xi)\Vert\teps^{\rm tr}\Vert^{2},\label{eq:Mori}
\end{equation}
where $C^{\rm int}$ is a positive constant depending only on material elastic constants. The interaction energy given by Eq. \eqref{eq:Mori} is maximal for $\xi=0.5$ and it symmetrically decreases towards $\xi=0$ and $\xi=1$ (see Fig. \ref{fig:uniaxial-fitting}b). As it is quadratic in $\xi$, the stress would decrease linearly with increasing $\xi$ due to this interaction term, which contradicts experimental observations. The reason of the symmetry of $E^{\rm int}$ with respect
to the $\xi\leftrightarrow(1-\xi)$ is given by equality of the energy of martensite inclusions with transformation strain $\teps^{\rm tr}$ within austenite matrix and (inversely) the energy of austenite inclusions without transformation strain within $\teps^{\rm tr}$-deformed martensite. According to experimental observations, this symmetry seems not fulfilled in NiTi, most likely due to nonlinear martensite behavior caused by its low
reorientation stress. Indeed, if we split $E^{\rm int}$ into two independent terms (notice that Eq. \eqref{eq:Mori} is recovered for $C_{\rm MA}^{\rm int}=C_{\rm AM}^{\rm int}=C^{\rm int}$):

\begin{equation}\label{eq:Mori-assym}
E^{\rm int}=C_{\rm MA}^{\rm int}\xi(1-\xi)^{2}\Vert\teps^{\rm tr}\Vert^{2}+C_{\rm AM}^{\rm int}\xi^{2}(1-\xi)\Vert\teps^{\rm tr}\Vert^{2},
\end{equation}
we are able to successfully fit strain-softening material response obtained in experiments. Figure \ref{fig:uniaxial-fitting}a) shows two uniaxial stress-strain relations extracted from dedicated experiments by \citet{HAL-KYR} and \citet{ALA-SIT-17} together with their best fits obtained by our constitutive model. Note the exceptionally good match in the (non-linear) strain-softening stage. Figure \ref{fig:uniaxial-fitting}b) reveals the asymmetric contributions of the first and second term on the right-hand side of the relation \eqref{eq:Mori-assym} to the total $E^{\rm int}$ responsible for the non-linear softening in E1 (the first term dominates). The resulting $E^{\rm int}$ in \eqref{eq:Mori-assym} can be simply considered as the energy of martensite inclusions (with transformation strain $\teps^{\rm tr}$) and austenite inclusions (without inelastic strain) within the average
austenite-martensite matrix with inelastic strain $\xi\teps^{\rm tr}$.

Finally, we can recast $E^{\rm int}$ into a form more suitable for model regularization and implementation into FEM:

\begin{equation}
\label{eq:Mori-regul}
E^{\rm int}=C_{\rm MA}^{\rm int}\xi\Vert\teps^{\rm tr}-(\xi\teps^{\rm tr})_\omega \Vert^{2}+C_{\rm AM}^{\rm int}(1-\xi)\Vert(\xi\teps^{\rm tr})_\omega\Vert^{2}.
\end{equation}
Here, variables $\xi$ and $\teps^{\rm tr}$ describe local material properties, whereas the product arrested in $(\cdot)_\omega$ representing average matrix inelastic strain is obtained by averaging inelastic strain within a certain neighborhood as specified below.

\section{Global formulation and mathematical properties}
\label{sec-maths}

In this section, we summarize the formulation of the extended model and give the main mathematical properties.

As already mentioned in Section \ref{sec-local}, we assume that the specimen occupies the domain $\Omega$ in the reference configuration, on which we define all variables. We will denote by $\partial \Omega$ the boundary of the reference configuration and assume
the following splitting $\partial \Omega = \Gamma_\mathrm{D} \cup
\Gamma_\mathrm{N}$ with $\Gamma_\mathrm{D}$,
$\Gamma_\mathrm{N}$ disjoint. On the part $\Gamma_\mathrm{N}$ the surface
force $F_\mathrm{surf}$ is acting on the specimen while on the
part $\Gamma_\mathrm{D}$ Dirichlet boundary conditions for the
displacement are prescribed. For simplicity, we restrict ourselves
here to \emph{zero Dirichlet boundary conditions}; i.e., $u(x) =
0$ on $\Gamma_\mathrm{D}$.

Moreover, while even the elastic part of the free energy \eqref{eq-energy} is not convex in $\xi$ nor $\varepsilon^\mathrm{tr}$, we shall, for mathematical considerations, rather switch variables and work with the volume fraction $\xi$ and the inelastic strain $\varepsilon^\mathrm{in}$ defined in \eqref{def-inelastic-strain}. In these variables then, at least the local part of the energy, is indeed convex. For the readers convenience, let us recall the definition of the energy and dissipation reformulated in the inelastic strain:
\begin{align}
f^{T}(\varepsilon,\xi,\varepsilon^\mathrm{in})&=\frac{1}{2}(\varepsilon- \varepsilon^\mathrm{in}):\mathbb{C}(\xi):(\varepsilon- \varepsilon^\mathrm{in})+\Delta s^{AM}(T-T_{0})\xi,
\label{eq-energy-inelastic} \\
d^T(\varepsilon^{\rm in},\xi, \dot{\varepsilon}^{\rm in},\dot{\xi})
  &=  \begin{cases}
  \Delta s^{AM}\left[T_0 - M_{\rm s} + \xi (M_{\rm s} - M_{\rm f})\right]|\dot{\xi}|
  +\; \sigma^{\rm reo}(T)\| \dot{\varepsilon}^\mathrm{in} \| & {\rm if }\; \dot{\xi} \geq 0, \\
  \Delta s^{AM}\left[A_{\rm f} - T_0 + \xi (A_{\rm s} - A_{\rm f})\right]|\dot{\xi}|
  +\; \sigma^{\rm reo}(T)\left[\|\frac{\dot{\xi}}{\xi}\varepsilon^{\rm in}\| +
  \| \dot{\varepsilon}^\mathrm{in}-\frac{\dot{\xi}}{\xi}\varepsilon^{\rm in}\|\right] & {\rm if}\; \dot{\xi} < 0.
  \end{cases}
\end{align}

Due to the non-local character of the newly added interaction energy term, we will work with the total energy given by
\begin{equation}\label{eq-def-energy-tot}
\mathcal{E}(t,u,\xi, \varepsilon^\mathrm{in}) := \int_\Omega
f^T(\varepsilon(u(x)),\xi(x), \varepsilon^\mathrm{in}(x)) + r_\mathrm{nonloc}(\xi(x), \varepsilon^\mathrm{in}) -
F_\mathrm{vol}(t)\cdot u \, \mathrm{d} x - \int_{\Gamma_\mathrm{N}}
F_\mathrm{surf}(t)\cdot u \, \mathrm{d} S,
\end{equation}
where $F_\mathrm{vol}$ and $F_\mathrm{surf}$ is the prescribed volume force acting on
the specimen. This force, as well as the surface force may depend on the time variable $t$. As we will assume that the evolution of the temperature in $T$ is a \emph{given} function of time, the dependence of the total energy on the temperature is captured again through the time variable.

Recall from section \ref{sec-nonloc} that energy contribution modelling localization is given by
$$
r_\mathrm{nonloc}(\xi(x), \varepsilon^{\rm in}(x)) = \begin{cases}
C_{\rm MA}^{\rm int} \frac{1}{\xi}\|\bm{\varepsilon}^{\rm in}(x)-\xi(x) (\bm{\varepsilon}^{\rm in})_\omega\|^2 + C_{\rm AM}^{\rm int}(1-\xi)\Vert(\varepsilon^{\rm in})_\omega\Vert^{2} & \text{if $\xi > 0$} \\
0  & \text{if $\xi = 0$}
\end{cases}
$$
and
\begin{equation} \label{eq-Gauss-regul}
(\bm{\varepsilon}^{\rm in})_\omega(x) = \frac{1}{\int_\Omega \mathcal{G}_\omega(x-y) \mathrm{d} y}\int_\Omega \bm{\varepsilon}^{\rm in}(y) \mathcal{G}_\omega(x-y) \mathrm{d} y,
\end{equation}
where $\mathcal{G}_\omega$ is a smooth function that models the averaging through the neighborhood. In particular, we assume that $0 \leq \mathcal{G}_\omega(\cdot) \leq 1$ and $\int_{\R^3} \mathcal{G}_\omega(x)\,\mathrm{d}x = 1$ and that $\mathcal{G}_\omega$ is smooth. The particular form of $\mathcal{G}_\omega$ is specified in Section \ref{sec-num}, but is not important from the mathematical point of view.

Let us also notice that we use the weighting factor $1/{\int_\Omega \mathcal{G}_\omega(x-y) \mathrm{d} y}$ in front of the averaging. In most cases, this factor will be just one, but it may play a role once the point $x$ is near or on the boundary. In this situation, we apply averaging only over the available specimen.

\begin{remark}
In fact, it is unclear how exactly the averaging should be designed near the boundary of the specimen $\Omega$. Here we choose to restrict the area of averaging to $\Omega$ \citep[as common in nonlocal models, see][]{PEE-GEE-01}, which essentially means that for points very near to the boundary the volume over which the averaging is performed gets smaller. Of course, from the physical point of view, this does not take into account that near the boundary the specimen experiences less geometric constrains and, e.g., the martensitic transformation may initiate easier. Nevertheless, capturing these effects is challenging from the modelling point of view and may have a little effect in practice. This is because the physically justified radius of averaging kernel should involve several neighboring grains forming the microstructure; in common NiTi components, this would be usually at the order of few microns, i.e. far below the reasonable mesh size.
\end{remark}

Similarly, we define the total dissipation which, however, has only local contributions.
\begin{equation}\label{eq-def-Diss-tot}
\mathcal{D}(t,\xi, \varepsilon^\mathrm{in}, \dot{\xi}, \dot{\varepsilon}^\mathrm{in}) := d^T(\varepsilon^{\rm in}(x),\xi(x), \dot{\varepsilon}^{\rm in}(x),\dot{\xi}(x)) \mathrm{d} x;
\end{equation}
notice that the overall dissipation depends explicitly on time which is caused by the fact that the constants in the dissipation function are dependent on the temperature, which may depend on time.

According to the generalized standard solids theory, the evolution of the specimen is given by balancing the conservative and dissipative force at all times $t \in [0,T]$, where $T$ is assumed to be the final time of the evolution. Formally, we may write that
\begin{equation}
\label{force-balance}
\underbrace{\partial_{(u,\xi, \varepsilon^\mathrm{in})}\mathcal{E}}_\text{conservative force}+\underbrace{\partial_{(\dot{\xi}, \dot{\varepsilon}^\mathrm{in})}\mathcal{D}}_\text{dissipative force} \ni 0 \qquad \text{for all $t \in [0,T]$},
\end{equation}
along with the constraint that the state variables remain in the admissible space $\mathcal{Q}:=\mathcal{U} \times \mathcal{V}$ with
\begin{align}
\mathcal{U} & := \{u \in {W}^{1,2}(\Omega,\mathbb{R}^3): u = 0 \text { on } \Gamma_\mathrm{D} \}, \\
\mathcal{V} & := \{(\teps^{\rm in}, \xi) \in
 {W}^{1,2}(\Omega,\mathbb{R}^{3\times3}) \times {W}^{1,2}(\Omega) \;:\;
                    \text{$\teps^{\rm in}$ is a traceless, symmetric matrix, } \nonumber \\
                &\qquad \langle \bm{\varepsilon}^{\rm in}(x)\rangle \leq \xi(x) \textrm{ for
                a.a.} x\in\Omega \textrm{ , and } 0\leq \xi(x) \leq 1 \textrm{ for a.a. }x\in\Omega\},
\end{align}
for all times $t \in [0,T]$. Here, ${W}^{1,2}(\Omega)$ denotes the Sobolev space of functions having square integrable derivatives.

Let us notice that \eqref{force-balance} is indeed just a formal expression, which, due to the non-local character, needs to be formulated in an integral form. It is beyond the scope of the present work to elaborate on the mathematical properties of \eqref{force-balance}, instead we shall concentrate on \emph{time-discrete approximations} of solutions of \eqref{force-balance} that will also be computed in the numerical part.

In the spirit of
\citep{francfort-mielke,mielke-theil} we design a
time-discretization of \eqref{force-balance}
via the backward Euler method. To be more specific, we introduce a partition of the time-interval $[0,T]$ via $0= t_0 \leq t_1 \leq \ldots t_{N(\tau)}=\mathcal{T}$ with
$\max_i(t_{i+1}-t_i) \leq \tau$ and $\tau$ being some small parameter. It is expected (even if we do not give a formal proof here) that solutions of the proposed time-discrete problems will approximate solutions of \eqref{force-balance}.

Then we call the
triple $(u_k,\varepsilon^{\rm in}_k,
\xi_k)\in \mathcal{Q}$ \emph{a time-discrete solution of
evolution at time $t_k$} to \eqref{force-balance}
 at time-level $k=1,
\ldots, N(\tau)$ if it solves
\begin{align}\label{RTIP}
&\text{Minimize }\text{$\mathcal{E}(t_k,u,\varepsilon^\mathrm{in},\xi) + \mathcal{D}(t_{k},\xi_{k-1},\teps^{\rm in}_{k-1},\xi-\xi_{k-1},\teps^{\rm
in}-\teps^{\rm in}_{k-1})$} \nonumber\\
&\text{subject to $(u,\varepsilon^\mathrm{in},\xi)\in \mathcal{Q}$} \tag{TIP}
\end{align}
with $(u_\tau^0,\varepsilon^\mathrm{in, 0},\xi^0) = (u_0,\varepsilon^\mathrm{in}_0,\xi_0)
\in \mathcal{Q}$ defined through the initial condition.

We call the minimization problem in \eqref{RTIP} the \emph{time-incremental problem}. It is physically well-motivated by the idea that upon a small change in the environment, i.e. during a small time-step, the studied system will try to find the relaxed state by minimizing the energy plus the dissipation needed to transit to the new state. In other words, the state variables describing the specimen will change if this yields a gain in the free energy larger that the dissipation.

The model proposed in this paper is well defined in the sense that \eqref{RTIP} possesses a solution. This is shown in the next proposition:

\begin{proposition}
\label{existence-TIP}
Let $(\xi_{k-1}\varepsilon^{\rm in}_{k-1}) \in L^\infty(\Omega) \times L^\infty(\Omega; \R^{3 \times 3})$ and that $F_\mathrm{vol} \in C^0(\Omega \times [0,T])$ as well as $F_\mathrm{surf} \in C^0(\Gamma_\mathrm{N} \times [0,T])$.
Then there exists
a triple $(u_k,\varepsilon^{\rm in}_k, \xi_k) \in \mathcal{Q}$ that solves \eqref{RTIP}.
\end{proposition}

In this proposition we, as is standard, denoted $L^\infty(\Omega)$ the
space of measurable functions that are bounded almost everywhere and $C^0(\cdot)$ stands for the space of continuous functions.

We postpone the proof of Proposition \ref{existence-TIP} to the appendix and make at this points only the following remark:

\begin{remark}[Convexity]
\label{rem-convexity}
Let us note that, thanks to the localization terms, the total energy $\mathcal{E}(t,u,\xi, \varepsilon^\mathrm{in})$ is not a convex function of its variables. This is best seen if we replaced the averaged term $(\varepsilon^\mathrm{in})_\omega$ by its local counterpart $\varepsilon^\mathrm{in}$ and rewrite (as in Section \ref{sec-nonloc}) $\varepsilon^\mathrm{in} = \xi \varepsilon^\mathrm{tr}$. Then, the localization energy would correspond to a double-well potential in $\xi$, which is clearly non-convex. This feature still persists if we allow for the averaged version $(\varepsilon^\mathrm{in})_\omega$. However, without this averaging, it could happen that very fine spatial oscillations between austenite and martensite appear because the interface between them is allowed to be infinitely (atomically) sharp. This is prohibited by the averaging term and allows to show existence of minimizers.

Nevertheless, let us note that due to the non-convexity one can expect several difficulties in calculations: solutions to \eqref{RTIP} are not necessarily unique and jumps in the temporal evolution of the variables may appear. This is related to the fact that, if the material would be completely homogeneous and there were no ``localization sites'', it could remain e.g. in austenite much longer than is energetically favourable and then abruptly transform at some random place \citep[cf.][]{ALE-BER-15}. However, such an ``indeterminacy in reponse'' could be easily removed by adding an initial imperfection to the numerical model: either local variation of geometry \citep[as in][]{SHA-KYR-97b,JIA-KYR-LAN-17} or local variation of material properties \citep[as in][]{ARM-BZ-14,SCIENCE}.

As was already noted in \citep{SED-FRO-IJOP}, apart from the non-local term, the energy function is
\emph{convex} in the its variables, which can be seen by simply
calculating the Hessian of $f^T$. Actually, one relies here on the fact that the
functions $h(x,y) = x^2/y$ and $\tilde h(x,y) =
x^2/(1-y)$ are convex, provided $0 \leq y \leq 1$. Moreover, the dissipation function
is \emph{convex} in the rate variables for thermodynamic consistency. This carries over to the discrete setting, as designed in \eqref{RTIP}.
\end{remark}

\section{Numerical simulations}
\label{sec-num}

In the following, we perform finite element simulations to illustrate the ability of the proposed constitutive model to capture the localization effect. The first simulation mimics the experimentally well-examined situation of a thin NiTi ribbon loaded in uniaxial tension \citep[e.g.][]{SHA-KYR-97,ZHA-SUN-10,JIA-KYR-17,BIA-PER-18}, the other one -- bending of a tube -- involves also compression of the material. In both of them, we employ generic material properties with pronounced strain softening in (uniaxial) tension and strain hardening in compression, see Fig. \ref{fig:uniaxial-fitting} and \ref{sec-append-B}, and presume spatially- and temporally-constant temperature corresponding to a quasistatic superelastic loading scenario. Before presenting the results, we sketch how the constitutive model can be incorporated into the finite element method.

\subsection{Implementation into the finite element method}

We implemented the constitutive model into the finite element software package \textit{Abaqus FEA} via User MATerial subroutine interface. The starting point is the time incremental problem \eqref{RTIP}, which can be divided into two minimization subproblems \citep{FRO-BEN-MMS}. The first one corresponds to finding displacement vector $u$ for fixed internal variables and prescribed boundary conditions (and temperature), which is a standard task for any finite element software package. In case of the local model (\ref{sec-local}), the second subproblem shrinks into searching for optimal values of internal variables in every single material point at given strain and temperature, which is handled by the UMAT subroutine as outlined in \citep{SED-FRO-IJOP}. The numerical solution of the complete problem \eqref{RTIP} is then resolved by an iterative procedure manged by \textit{Abaqus}, see \citep{SED-FRO-IJOP, FRO-BEN-MMS} for details.
In the nonlocal model, the evolution of internal variables in a material point inherently depends on the response of neighboring material points, hence, the individual treatment is not possible anymore.

A possible way how to implement the nonlocal integral regularization formulated in \eqref{eq-Gauss-regul} was proposed by \citet{BOB-TEJ-04,BOB-TEJ-05}. In their approach, a virtual mesh of the computational domain is added (so that nodes of the virtual mesh coincide with those of the original mesh of the body) and nonlocal variables are treated within this mesh. As described in our previous work \citep{SMST-LOC}, it is possible to complement the UMAT subroutine by subroutine UEXTERNALDB, which is activated after each increment in order to update the values of the nonlocal variable and pass them back to the UMAT. The following numerical simplifications allow to reduce computation costs \citep{SMST-LOC}: (i) the integration kernel has a finite spatial coverage \citep[cf.][]{PEE-GEE-01,JIR-ROL-03,BOB-TEJ-04}; (ii) in each time increment, the nonlocal variable is computed based on values of internal variables obtained in the previous converged increment (staggered computation method), (iii) only the scalar variable $\xi$ is averaged in the superelastic loading regime. Although there are more candidates for the averaging function $\mathcal{G}_\omega$ appearing in \eqref{eq-Gauss-regul}, homogeneous and isotropic ones are usually preferred in the literature mainly for practical reasons \citep{PEE-GEE-01,JIR-ROL-03}; a typical example is the (three-dimensional) Gauss distribution defined as
\begin{equation}\label{eq:Gauss-kernel}
\mathcal{G}_\omega(x-y) := \frac{1}{(2\pi)^{3/2}\omega^3}\exp{\left[-\frac{\|x - y\|^2}{2\omega^2}\right]}.
\end{equation}

Let us finally note that a conceptually analogous constitutive model was successfully implemented and employed for a simulation of propagation of the martensite band front in a thin NiTi wire under tension in our previous work \citep{SCIENCE}.

\subsection{Tension of a NiTi ribbon}

\begin{figure}
\includegraphics[scale=0.45]{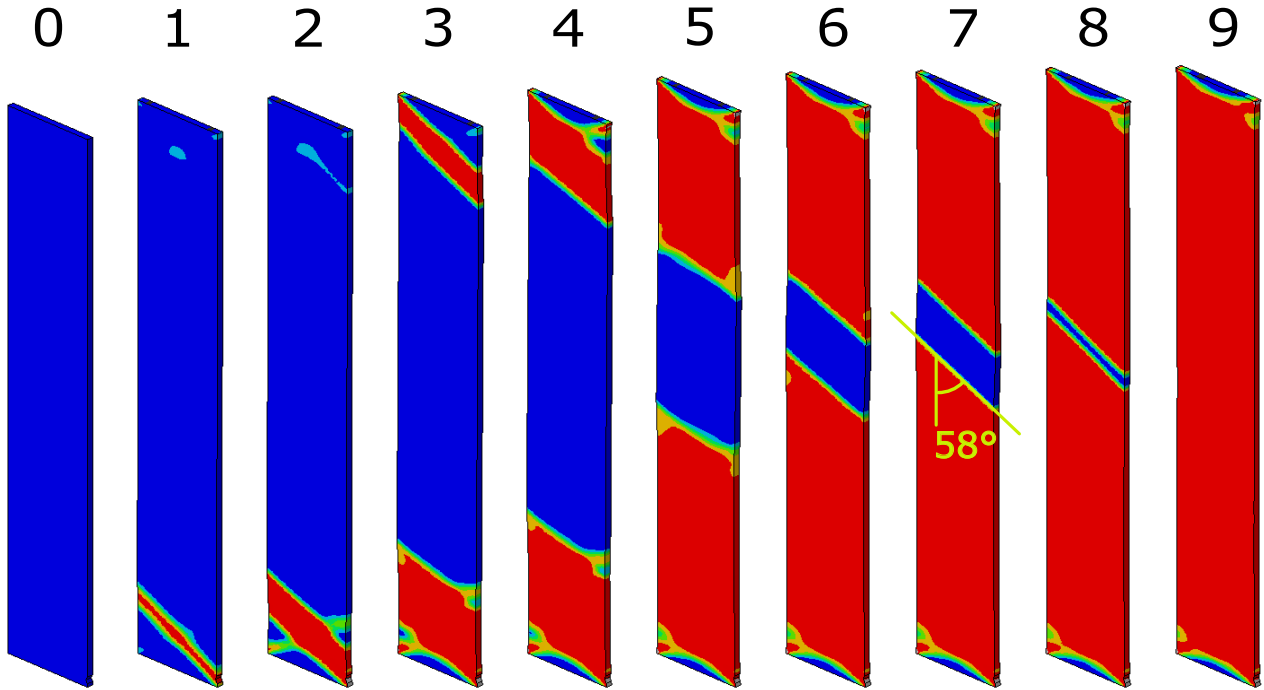}
\caption{Snapshots from a three-dimensional simulation of the NiTi ribbon in tension with distribution of volume fraction of martensite in color (for legend see the VFM colorbar in Figure \ref{fig:tube}). The average angle of inclination of the planar martensite band front with respect to the axis of the ribbon is marked in \textcircled{{7}}.} \label{fig:ribbon}
\end{figure}


We use a geometric model of a NiTi ribbon in the form of a rectangular prism with dimensions of the base 15\,mm and 1\,mm and with 120\,mm in length. The body is partitioned into a uniform mesh of 3,200 ($20 \times 1 \times 160$) identical hexaedral (brick) elements with linear interpolation (C3D8) to reduce any directional bias. For an easy initiation of the localization pattern, a small geometric imperfection in the form of a V-shaped indent is imposed on one lateral side of the strip so that the width of the most reduced cross-section is 14.7\,mm. The indent is located 1.5\,mm from one of the bases.

The ribbon is loaded axially by prescribing displacement boundary conditions at both bases; all other surfaces are stress free. All displacement degrees of freedom are fixed at the base closer to the indent. At the other one, the axial displacement is incrementally prescribed so that the maximum displacement-to-initial length ratio is $7.5\%$, whereas both lateral displacements are fixed.

Figure \ref{fig:ribbon} presents several snapshots of the distribution of volume fraction of martensite within the ribbon during loading. The common features of this type of localization patterns can be observed: nucleation event in the form of an inclined thin martensite band crossing the sample \textcircled{\raisebox{-0.9pt}{1}}, restoring the momentum balance by nucleation of additional bands with either the opposite inclination or from the other end of the ribbon \textcircled{\raisebox{-0.9pt}{2}}--\textcircled{\raisebox{-0.9pt}{4}}, propagation of bands along the sample either via movement of one inclined phase interface or
via alternating between the two of them producing the crisscross (''finger-like'') pattern
\textcircled{\raisebox{-0.9pt}{4}}--\textcircled{\raisebox{-0.9pt}{8}}, and their coalescence at the final stage of loading \textcircled{\raisebox{-0.9pt}{9}}. The average angle of inclination of the matensite band front (obtained from several snapshots of the simulation) is $(58\pm 2)^\circ$, which is close both to experimental and modeling results obtained in the literature \citep[cf.][]{SHA-KYR-97b,AZA-RAJ-07,GRO-WAG-10,JIA-KYR-17,MRH-STU-19} and not far from the idealized theoretical \emph{two-dimensional} analysis (e.g. in \citep{SHA-KYR-97b}) giving the value $54.74^\circ$.

\subsection{Bending of a NiTi tube}

\begin{figure}
\includegraphics[scale=0.23]{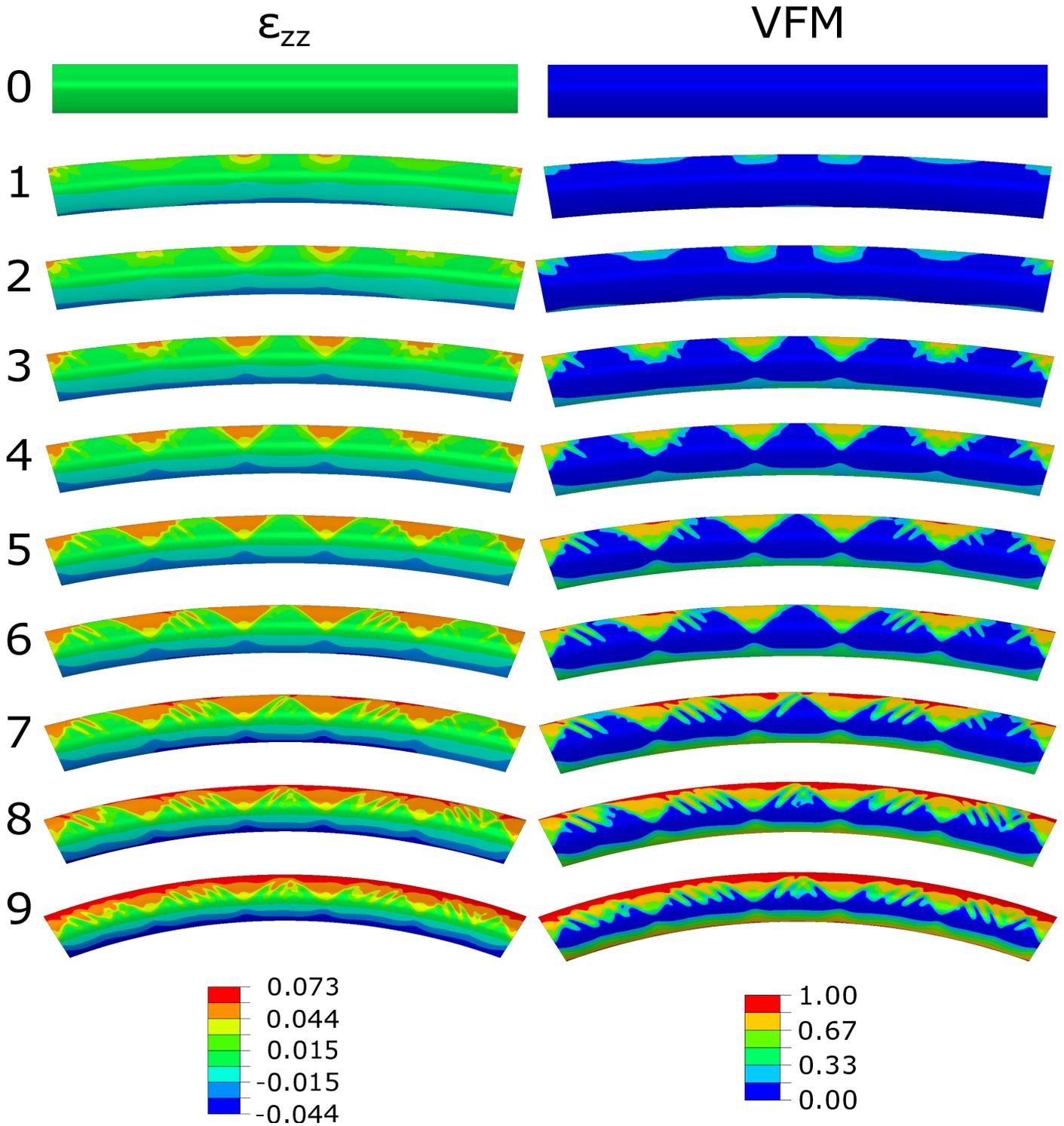}
\caption{Snapshots from a three-dimensional simulation of the NiTi tube in bending. The distribution of the axial component of strain on the left, the distribution of the volume fraction of martensite on the right.}\label{fig:tube}
\end{figure}

We consider an ideal tube of (outer) diameter 3.5\,mm, wall thickness 0.25\,mm and length 25\,mm partitioned into a uniform mesh of hexaedral elements (C3D8R) with four elements through the thickness, 96 along the circumference and 250 along the length (96,000 elements in total), i.e. full 3D geometry is modelled.
Displacement boundary conditions are imposed on both bases (annuli) of the tube, remaining surfaces of the tube are stress free. Utilizing the *COUPLING *KINETIC feature of the Abaqus CAE software, mutually inverse rotation of both annular surfaces around an axis perpendicular to the tube axis is prescribed ($\pm 25^\circ$), which leads to the desired bending of the whole tube, and rigid body motion is excluded.

The simulation is motivated by dedicated experiments published in \citep{BEC-KYR-14,REE-DAL-14, WAT-REE-18,JIA-KYR-LAN-17}. Just as in the previous section, we focus on demonstrating the capability of the proposed model to capture the key features of the behavior rather than attempting to reproduce the experiments quantitatively. However, with respect to the diameter to thickness ratio and to the boundary conditions, we are closer to the work of \citet{WAT-REE-18}.

In Fig. \ref{fig:tube} we present both strain and phase distributions to emphasize the different values of maximum transformation strains linked with full martensite in tension and compression, cf. Fig. \ref{fig:uniaxial-fitting}a). The $\varepsilon_{zz}$ strain is the diagonal component of the total strain tensor in the direction of the symmetry axis of the tube in the reference configuration. Again, one can observe several common features of the localization patterning: appearance of inclined finger-like structures of high strain \textcircled{\raisebox{-0.9pt}{4}},\textcircled{\raisebox{-0.9pt}{5}}, their growth and multiplication \textcircled{\raisebox{-0.9pt}{6}},\textcircled{\raisebox{-0.9pt}{7}}, and their crisscrossing and coalescence into wedge-shaped regions \textcircled{\raisebox{-0.9pt}{8}},\textcircled{\raisebox{-0.9pt}{9}} covering most of the part of surface undergoing predominantly tensile straining. The shift of the neutral axis effectively increasing this part of the surface can be observed in accordance with experimental observations in \citep{WAT-REE-18, JIA-KYR-LAN-17}. The tension-compression asymmetry embodied in the constitutive law leads to much lower (absolute) values of strain on the lower part of the tube, and, foremost, to the homogeneous distribution of strain (and VFM) there. For comparison, we performed an identical simulation except for a modified constitutive law incorporating strain-softening also in compression. In that case, the wedge-like localization pattern was observed (not presented here) also on the lower part of the tube (where compression loading dominates) \citep[also cf.][]{JIA-KYR-LAN-17}.

\section{Discussion and Conclusions}

We extended a well-established constitutive model tailored for NiTi-based shape memory alloys by an interaction energy term allowing to capture the localization effects. The summary of  experimental observations in Section \ref{sec-nonloc} provided hints for searching a suitable form of such an energy term within the classical Mori-Tanaka approach. The key features of the final form given by Eq. \eqref{eq:Mori-assym} are: (i) it is a sum of two independent terms, (ii) both terms are non-convex in $\xi$, (iii) both terms exhibit quadratic dependence on $\|\bm{\varepsilon}^{\rm tr}\|$.

\begin{enumerate}[(i)]

\item
Thanks to two independent constants appearing in the interaction energy, we gained more freedom to adjust the strain-softening constitutive law to available experimental measurements, as illustrated in Fig. \ref{fig:uniaxial-fitting}; particularly, we were not limited to the solely linear decrease of stress with strain.

\item
Adopting the localization contribution energy $E^{\rm int}$ means that the overall energy becomes non-convex. This is, of course, a desired effect because in this way it prefers the pure states of austenite as well as martensite, hence the material transforms in a localized rather than a homogeneous way. Nevertheless, non-convex energy contributions always present possible difficulties from the mathematical as well as the computational point of view.

First, it may happen that, because of the non-convexity, solutions to the time-incremental problems \eqref{RTIP} do not even exist. This is typically caused by the appearance spatial oscillations that tend to be infinitely fine. Indeed, it could happen that the material would develop ``infinitely thin" stripes of martensite as well as austenite, which would effectively prohibit the existence of solutions. In numerical implementations, the width of such oscillations would be given just by the mesh-size; this holds true even in the case one interface would develop.
We exclude such pathological behavior by averaging in some variables in the non-local term, which introduces a final width of the austenite-martensite interface. 

While introducing non-locality allows us to exclude possible undesired effects, this reflects in more costly calculations. Nonetheless, in many situations, it is justified to replace the non-local averaged quantities by local ones. Particularly, in the three-dimensional setting, for a large class of loading regimes, an infinitely fine austenite-martensite interface cannot be formed just by geometric reasons leading naturally to the presence of a phase gradient region between the two phases.
\footnote{Hence, in practice, it may be enough to do all calculations using just the local variables and checking at the same time the size of the gradient of the volume fraction of martensite. If this does not get too large, it is justified to perform the calculations in local variables only, which reduces the computational cost.}

Finally, let us notice that independently of whether we use the averaging kernels or not, non-convexity of the problem is connected with the appearance of multiple local minimizers in \eqref{RTIP}. As, in numerical implementations, we can always find only local minimizers, this may lead to non-uniqueness of the results. A typical situation is that in calculations the austenite-martensite transformation happens too late. To avoid this, an effective strategy is to add e.g. stress concentrators into the model.

Let us note that the above observations are also valid if we did not solve the minimization problem \eqref{RTIP} but instead the associated variational inequalities.

\item
The factor $\Vert\boldsymbol{\boldsymbol{\varepsilon}}^{\rm tr}\Vert^{2}$ in the interaction energy brings the loading-mode dependency into the model: when transformation strain for a particular loading direction is large, the interaction term dominates over the loading-mode insensitive ``transformation hardening'' defined in the model by the difference of transformation temperatures $(M_{s}-M_{f})$ and $(A_{s}-A_{f})$. If transformation strain in tension is almost twice the transformation strain in compression (as suggested by experiments), the interaction energy is almost four times larger in tension than in compression; this results in strain-softening response in tension and hardening-like in compression.

This observation emphasizes the need for correct description of the so-called \emph{transformation strain surface} (i.e. the boundary of the set of all available transformation strain tensors), which is defined by a 1-homogeneous convex function in our model, see the constraint in \eqref{eq:eps-tr-def} and \ref{sec-append-B}. Our particular form of the transformation strain surface allows to capture tension/compression asymmetry of the maximum transformation strain (by involving the third invariant of the strain tensor), but does not include possible material anisotropy. The question of a suitable form of the transformation strain surface appropriate for simulations of NiTi components still remains open \citep{APPA-ANISO}, and several experimental works suggest different forms involving also material anisotropy given by usually strong $\langle 111 \rangle$ texture in NiTi drawn components \citep{REE-DAL-12, BEC-KYR-16}.

Involving effects of material texture and anisotropy by redefining transformation strain surface could be done straightforwardly in our model as we only assumed that the surface is described by a 1-homogeneous convex function. However, it is questionable whether all experimentally observed effects of material anisotropy on localization of martensitic transformation can be covered solely by a proper description of the transformation strain surface. For example, microstructure observations of martensitic transformation on NiTi twisted tube by \citet{PEN-FAN-08} revealed that martensite appears in the form of almost parallel lamellae with inclination about $26.5^\circ$ from the axial direction of the tube. Such a strong preferential orientation of austenite-martensite habit planes can be explained by $\langle 111 \rangle$ material texture. By calculating all possible habit plane orientations  -- 24 Type II habit plane variants from \citep{MAT-OTS-87} were considered -- using the mathematical theory of martensitic microstructures by \citet{Ball1}, it can be shown that there are several possible habit planes nearly parallel to the $[111]_{\rm A}$ direction (there are at least nine different habit planes inclined less than $10^\circ$ w.r.t. $[111]_{\rm A}$ direction), while there are no habit planes nearly perpendicular to this direction (the normal to the least  favourable one is inclined just in an angle of $16.5^\circ$ w.r.t. $[111]_{\rm A}$). Among the nine habit planes nearly parallel to $[111]_{\rm A}$, three possible orientations correspond to remarkably large resolved shear strains. Hence, it is plausible that the first nuclei of stress-induced martensite can be encapsulated by such habit planes under the applied twisting. Such a strong orientation of forming microstructure could alter both expression of the interaction energy (derived in the Section \ref{sec-nonloc} with the assumption of spherical inclusions) but also the way of calculation of non-local variables: the isotropic integration kernel in Eq. \eqref{eq:Gauss-kernel} could be replaced by an anisotropic one to reflect directional dependence of interaction between inclusions. We plan to address these issues in future work.

\end{enumerate}

Finally, in addition to the above discussed key features of the definition \eqref{eq:Mori-assym}, its motivation stemming from the Mori-Tanaka method also provided a hint for the physically plausible regularization of the non-convex model expressed by \eqref{eq:Mori-regul}. Then, we could proceed by performing its basic mathematical analysis, which provided a sound basis for a numerical implementation of the time-discretized problem \ref{RTIP} into FEM excluding pathological behavior in simulations.

The simulation of a NiTi tube subjected to bending showed that the model is able to capture the difference between loading in tension -- where characteristic patterns of localized deformation appeared -- and compression -- where homogeneous deformation persists. In our previous work \citep{SMS-SNAKE}, the model was also employed in a study on bending of a NiTi wire structure. Thanks to direct comparison of simulations with x-ray microdiffraction data, it was confirmed that localization strongly affects the mechanical response also in such a case.

The computed localization patterns on the bent tube as well as those on the NiTi ribbon under tension exhibited some features commonly observed in experiments. As pointed out by many authors \citep[e.g.][]{SIT-LIU, GRO-WAG-10,XIA-LEI-17,ZHA-HE-18}, particular geometrical forms and propagation modes of the localization bands -- even for the specific type of sample geometry and loading mode -- strongly depend on the material (composition, processing), dimensions of the sample, and boundary conditions.\footnote{And, beyond the quasistatic approximation, they also depend on the deformation rate \citep{ZHA-SUN-10}.} Thus, the presented FE model also provides a powerful tool for further exploration of these issues.

\section*{Acknowledgement}
We acknowledge the financial support of the Czech Science Foundation via project No. GA18-03834S and of the Ministry of Education, Youth and Sports via project \\
No. CZ.02.1.01/0.0/0.0/16\_013/0001794 (European Spallation Source - participation of the Czech Republic – OP) funded within the Operational Programme Research, Development and Education. We would like to thank Dr. E. Alarcon for providing us the experimental data used in Fig. \ref{fig:uniaxial-fitting}a).

\appendix
\section{Mathematical analysis}

In this section, we prove Proposition \ref{existence-TIP} by the direct method of calculus of variations (cf. e.g. \citep{dacorogna}). To do so, we rely on the convexity of the minimized function in all term except the localization one, in the latter we rely on the used averaging. In fact, we first need to realize that the localization contribution to the energy is continuous, i.e. that $r_\mathrm{nonloc}(\xi, \bm{\varepsilon}^{\rm in})$ is a continuous function on $[0,1] \times \mathbb{R}^{3 \times 3}$. The only thing we have to verify is that
$$
\lim_{\xi \to 0} \frac{\|\bm{\varepsilon}^{\rm in}-\xi (\bm{\varepsilon}^{\rm in})_\omega\|^2}{\xi} = 0.
$$
We use that $\langle \cdot \rangle$ is an equivalent norm on $\mathbb{R}^{3 \times 3}$ meaning that there exists a constant c such that $\|\varepsilon\| \leq c \langle \varepsilon \rangle$ for some constant $c$ and all $\varepsilon \in \mathbb{R}^{3 \times 3}$. Owing to this
$$
0 \leq \frac{\|\bm{\varepsilon}^{\rm in}-\xi (\bm{\varepsilon}^{\rm in})_\omega\|^2}{\xi} \leq c \frac{\langle\bm{\varepsilon}^{\rm in}-\xi (\bm{\varepsilon}^{\rm in})_\omega \rangle^2 }{\xi} \leq c \frac{\xi^2+\xi^2 \langle (\bm{\varepsilon}^{\rm in})_\omega \rangle^2}{\xi} \stackrel{\xi \to 0}{\longrightarrow} 0,
$$
where we also used that $\langle \bm{\varepsilon}^{\rm in}\rangle \leq \xi$.
Furthermore, $r_\mathrm{nonloc}(\xi, \varepsilon^{\rm in})$ is convex function in $\varepsilon^{\rm in}, \xi)$, if we regard $\varepsilon^{\rm in})_\omega$ as a fixed independent variable. This follows from the positive definiteness of the Hessian and is related to the fact that $\frac{x^2}{y}$ is a convex function for $x,y > 0$.

Let us now turn to proving existence of minimizers to \eqref{RTIP}. We find a minimizing sequence $[u^j,[
\varepsilon^{\rm in}]^j,\xi^j)_{j\in\mathbb{N}} \subset \mathcal{Q}$ so that (for $j\to\infty$)
$$
\mathcal{E}(t_k,u^j,[\varepsilon^\mathrm{in}]^j,\xi^j) + \mathcal{D}(t_{k},\xi_{k-1},\teps^{\rm in}_{k-1},\xi^j-\xi_{k-1},[\varepsilon^{\rm
in}]^j-\varepsilon^{\rm in}_{k-1}) \to \mathrm{inf}.
$$
Due to the quadratic growth of the energy in the elastic part as well as due to the bounds imposed in $\mathcal{Q}$, we can assume that (at least for a subsequence denoted by the same indices) we have the following convergence results for $j\to\infty$:
\begin{align*}
    u^j & \rightharpoonup u_k \qquad \text{in } W^{1,2}(\Omega; \mathbb{R}^3), \\
[\varepsilon^{\rm in}]^j & \stackrel{*}{\rightharpoonup} \varepsilon^{\rm in}_k\qquad  \text{in } L^\infty(\Omega; \mathbb{R}^{3 \times 3}), \\
\xi^j & \stackrel{*}{\rightharpoonup} \xi_k \qquad \text{in } L^\infty(\Omega).\\
\end{align*}

We now show that $(u_k,\varepsilon^{\rm in}_k,\xi_k,) \in W^{1,2}(\Omega; \mathbb{R}^{3}) \times L^\infty(\Omega; \mathbb{R}^{3 \times 3}) \times  L^\infty(\Omega)$ belongs to $\mathcal{Q}$ is a solution of \eqref{RTIP}. Due to the convexity of the constraints, we know that
$$\langle \varepsilon^{\rm in}_k(x)\rangle \leq \xi_k(x),\quad \text{and} \quad 0 \leq \xi_k(x) \leq 1,\quad \text{a.e. on $\Omega$.}$$
Moreover, due to the convexity of the local part of the energy as well as the dissipation (in the rate variable) we see that
\begin{align*}
\int_\Omega
& f^T(\varepsilon(u_k(x)),\xi_k(x), \varepsilon^\mathrm{in}_k(x))  -
F_\mathrm{vol}(t_k)\cdot u_k + d^{T_k}(\xi_{k-1},\varepsilon^{\rm in}_{k-1},\xi_k-\xi_{k-1},\varepsilon^{\rm in}_k-\teps^{\rm
in}_{k-1}) \mathrm{d} x \\ &\qquad \qquad \qquad  - \int_{\Gamma_\mathrm{N}}
F_\mathrm{surf}(t_k)\cdot u_k \, \mathrm{d} S \\
&\quad \leq \liminf_{j \to \infty}
f^T(\varepsilon(u^j(x)),\xi^j(x), [\varepsilon^\mathrm{in}]^j(x))  -
F_\mathrm{vol}(t_k)\cdot u^j + d^{T_k}(\xi_{k-1},\varepsilon^{\rm in}_{k-1},\xi^j-\xi_{k-1},[\varepsilon^{\rm in}]^j - \teps^{\rm in}_{k-1}) \mathrm{d} x \\ &\quad \qquad \qquad \qquad  - \int_{\Gamma_\mathrm{N}}
F_\mathrm{surf}(t_k)\cdot u^j \, \mathrm{d} S
\end{align*}

So we only need to look at the nonlocal part. To this end, we realize that, for any fixed $\omega > 0$, we have that, for all $x \in \Omega$
$$
([\bm{\varepsilon}^{\rm in}]^j)_\omega(x) = \int_\Omega [\bm{\varepsilon}^{\rm in}]^j(y) \mathcal{G}_\omega(x-y) \mathrm{d} y \to \int_\Omega \bm{\varepsilon}^{\rm in}_k(y) \mathcal{G}_\omega(x-y) \mathrm{d} y = (\bm{\varepsilon}^{\rm in}_k)_\omega(x),
$$
so that $([\bm{\varepsilon}^{\rm in}]^j)_\omega \to (\bm{\varepsilon}^{\rm in})_\omega$ pointwise  in $\Omega$. Furthermore, since $([\bm{\varepsilon}^{\rm in}]^j)_\omega(x)$ is uniformly bounded, this extends to strong convergence in $L^p(\Omega; \mathbb{R}^{3 \times 3})$ for all $p \in [1,\infty)$. For the first term in the localization energy we use the rewriting
\begin{align*}
&\frac{\|[\varepsilon^{\rm in}]^j-\xi^j ([\varepsilon^{\rm in}]^j)_\omega\|^2}{\xi^j} = \frac{\|[\varepsilon^{\rm in}]^j-\xi^j ([\varepsilon^{\rm in}]^j)_\omega-\xi^j (\varepsilon^{\rm in}_k)_\omega+\xi^j (\varepsilon^{\rm in}_k)_\omega\|^2}{\xi^j}  \\ &\qquad = \frac{\|[\varepsilon^{\rm in}]^j-\xi^j (\varepsilon^{\rm in}_k)_\omega\|^2}{\xi^j} + 2\big([\varepsilon^{\rm in}]^j-\xi^j (\varepsilon^{\rm in}_k)_\omega\big)\cdot\big(([\varepsilon^{\rm in}]^j)_\omega- (\varepsilon^{\rm in}_k)_\omega\big)+ \xi^j \|([\varepsilon^{\rm in}]^j)_\omega- (\varepsilon^{\rm in}_k)_\omega\|^2,
\end{align*}
we see that the first term in the second line is convex in $(\xi_j,[\varepsilon^{\rm in}]^j)$, so that its integral is weakly lower semicontinuous and the integral of the latter two converges to 0 as $j \to \infty$. For the second term in the localization energy, we only need to look at
$$
\xi^j\Vert([\varepsilon^{\rm in}]^j)_\omega\Vert^{2}
= \xi^j\Vert(\varepsilon^{\rm in}_k)_\omega\Vert^{2} - 2 \xi^j (\varepsilon^{\rm in}_k)_\omega  \cdot\big(([\varepsilon^{\rm in}]^j)_\omega- (\varepsilon^{\rm in}_k)_\omega\big)+ \xi^j \|([\varepsilon^{\rm in}]^j)_\omega- (\varepsilon^{\rm in}_k)_\omega\|^2,
$$
where again the the first term on the right hand side is even linear in $\xi^j$ while the other two tend to zero as $j \to \infty$.

Combining the arguments above implies that $(u_k,\varepsilon^{\rm in}_k,\xi_k) \in W^{1,2}(\Omega; \mathbb{R}^{3}) \times L^\infty(\Omega; \mathbb{R}^{3 \times 3}) \times  L^\infty(\Omega)$ is a solution of \eqref{RTIP}

\section{Material parameters}
\label{sec-append-B}

The material parameters obtained as the best fit of experimental data E1 \citep{ALA-SIT-17} in Fig. \ref{fig:uniaxial-fitting} are summarized in Tab. \ref{tab:param}.

\begin{table}[]
\caption{\label{tab:param}Table of material parameters used in computations.}
\begin{center} 
\renewcommand{\arraystretch}{1.05}
{\small \begin{tabular}{@{}lllp{8.5cm}}
& & & \\ 
\hline
Parameter & Value & Unit & Brief description\\
\hline \\[-4.5mm]
$K$ & $148$ & {[}GPa{]} & Bulk modulus common to both phases.\\
$G^{\rm A},G^{\rm M}$ & $25, 15$ & {[}GPa{]} & Shear moduli of austenite and martensite.\\
$k$ & $0.072$ & {[}1{]} & Maximum transformation strain in tension.\\
$a$ & $0.99$ & {[}1{]} & Tension-compression asymmetry parameter. \\
$A_{\rm s},A_{\rm f}$ & $-30, -18$ & {[$^\circ$C]} & Martensite to austenite transformation temperatures. \\
$M_{\rm s},M_{\rm f}$ & $-37, -47$ & {[$^\circ$C]}  & Austenite to martensite transformation temperatures.\\
$T_0$ & $-35$ & {[$^\circ$C]} & Equilibrium austenite-martensite temperature. \\
$\sigma^{\rm reo}$ & $85$ & {[}MPa{]} & Martensite reorientation stress.\\
$\Delta s^{\rm AM}$ & $0.36$ & {[}MPa/$^\circ$C{]} & Difference between specific entropies of martensite and austenite.\\
$C^{\rm int}_{\rm MA}$ & $80$ & {[}MPa{]} & Parameter of the interaction energy in Eq. \eqref{eq:Mori-regul}.\\
$C^{\rm int}_{\rm AM}$ & $29$  & {[}MPa{]} & Parameter of the interaction energy in Eq. \eqref{eq:Mori-regul}.\\
\hline
\end{tabular}}
\end{center}
\end{table}

The particular form of the function confining the transformation strain $\langle \cdot \rangle$ from constraints \eqref{eq:eps-tr-def} and defining the transformation strain surface is \citep[cf.][]{SED-FRO-IJOP, APPA-ANISO}:
\begin{equation}
\langle \bm{\varepsilon}^{\rm tr} \rangle =
\frac{I_2(\bm{\varepsilon}^{\rm tr})}{k}
\frac{\cos\left(\frac{1}{3}{\arccos(1-a(I_3(\bm{\varepsilon}^{\rm
tr}) + 1))}\right)}{\cos\left(\frac{1}{3}{\arccos(1-2a)}\right)},
\label{eq:asymfc}
\end{equation}
where
\begin{equation}
I_2(\bm{x}) := \sqrt{\frac{2}{3}x_{ij}x_{ij}}, \quad I_3(\bm{x})
:= 4\frac{\det(\bm{x})}{(I_2(\bm{x}))^3}.
\end{equation}

\section*{References}

\bibliographystyle{elsarticle-harv}
\bibliography{liter-2019}

\begin{thebibliography}{71}
\expandafter\ifx\csname natexlab\endcsname\relax\def\natexlab#1{#1}\fi
\providecommand{\url}[1]{\texttt{#1}}
\providecommand{\href}[2]{#2}
\providecommand{\path}[1]{#1}
\providecommand{\DOIprefix}{doi:}
\providecommand{\ArXivprefix}{arXiv:}
\providecommand{\URLprefix}{URL: }
\providecommand{\Pubmedprefix}{pmid:}
\providecommand{\doi}[1]{\href{http://dx.doi.org/#1}{\path{#1}}}
\providecommand{\Pubmed}[1]{\href{pmid:#1}{\path{#1}}}
\providecommand{\bibinfo}[2]{#2}
\ifx\xfnm\relax \def\xfnm[#1]{\unskip,\space#1}\fi
\bibitem[{Ahmadian et~al.(2015)Ahmadian, Ardakani and Mohammadi}]{AHM-HOS}
\bibinfo{author}{Ahmadian, H.}, \bibinfo{author}{Ardakani, S.H.},
  \bibinfo{author}{Mohammadi, S.}, \bibinfo{year}{2015}.
\newblock \bibinfo{title}{Strain-rate sensitivity of unstable localized phase
  transformation phenomenon in shape memory alloys using a non-local model}.
\newblock \bibinfo{journal}{Int. J. Solids Struct.} \bibinfo{volume}{63},
  \bibinfo{pages}{167--183}.
\bibitem[{Alarcon et~al.(2017)Alarcon, Heller, Chirani, {\v{S}}ittner,
  Kope{\v{c}}ek, Saint-Sulpice and Calloch}]{ALA-SIT-17}
\bibinfo{author}{Alarcon, E.}, \bibinfo{author}{Heller, L.},
  \bibinfo{author}{Chirani, S.}, \bibinfo{author}{{\v{S}}ittner, P.},
  \bibinfo{author}{Kope{\v{c}}ek, J.}, \bibinfo{author}{Saint-Sulpice, L.},
  \bibinfo{author}{Calloch, S.}, \bibinfo{year}{2017}.
\newblock \bibinfo{title}{Fatigue performance of superelastic {N}i{T}i near
  stress-induced martensitic transformation}.
\newblock \bibinfo{journal}{Int. J. Fatigue} \bibinfo{volume}{95},
  \bibinfo{pages}{76--89}.
\bibitem[{Alessi and Bernardini(2015)}]{ALE-BER-15}
\bibinfo{author}{Alessi, R.}, \bibinfo{author}{Bernardini, D.},
  \bibinfo{year}{2015}.
\newblock \bibinfo{title}{Analysis of localization phenomena in shape memory
  alloys bars by a variational approach}.
\newblock \bibinfo{journal}{Int. J. Solids Struct.} \bibinfo{volume}{73-74},
  \bibinfo{pages}{113--133}.
\bibitem[{Armattoe et~al.(2014)Armattoe, Haboussi and Ben~Zineb}]{ARM-BZ-14}
\bibinfo{author}{Armattoe, K.}, \bibinfo{author}{Haboussi, M.},
  \bibinfo{author}{Ben~Zineb, T.}, \bibinfo{year}{2014}.
\newblock \bibinfo{title}{A {2D} finite element based on a nonlocal
  constitutive model describing localization and propagation of phase
  transformation in shape memory alloy thin structures}.
\newblock \bibinfo{journal}{Int. J. Solids Struct.} \bibinfo{volume}{51},
  \bibinfo{pages}{1208--1220}.
\bibitem[{Auricchio et~al.(2014)Auricchio, Bonetti, Scalet and
  Ubertini}]{AUR-BON-14}
\bibinfo{author}{Auricchio, F.}, \bibinfo{author}{Bonetti, E.},
  \bibinfo{author}{Scalet, G.}, \bibinfo{author}{Ubertini, F.},
  \bibinfo{year}{2014}.
\newblock \bibinfo{title}{Theoretical and numerical modeling of shape memory
  alloys accounting for multiple phase transformations and martensite
  reorientation}.
\newblock \bibinfo{journal}{Int. J. Plast.} \bibinfo{volume}{59},
  \bibinfo{pages}{30--54}.
\bibitem[{Azadi et~al.(2007)Azadi, Rajapakse and Maijer}]{AZA-RAJ-07}
\bibinfo{author}{Azadi, B.}, \bibinfo{author}{Rajapakse, R.D.},
  \bibinfo{author}{Maijer, D.M.}, \bibinfo{year}{2007}.
\newblock \bibinfo{title}{Multi-dimensional constitutive modeling of {SMA}
  during unstable pseudoelastic behavior}.
\newblock \bibinfo{journal}{Int. J. Solids Struct.} \bibinfo{volume}{44},
  \bibinfo{pages}{6473--6490}.
\bibitem[{Badnava et~al.(2014)Badnava, Kadkhodaei and Mashayekhi}]{BAD-KAD}
\bibinfo{author}{Badnava, H.}, \bibinfo{author}{Kadkhodaei, M.},
  \bibinfo{author}{Mashayekhi, M.}, \bibinfo{year}{2014}.
\newblock \bibinfo{title}{A non-local implicit gradient-enhanced model for
  unstable behaviors of pseudoelastic shape memory alloys in tensile loading.}
\newblock \bibinfo{journal}{Int. J. Solids Struct.} \bibinfo{volume}{51},
  \bibinfo{pages}{4015--4025}.
\bibitem[{Ball and James(1987)}]{Ball1}
\bibinfo{author}{Ball, J.}, \bibinfo{author}{James, R.}, \bibinfo{year}{1987}.
\newblock \bibinfo{title}{{Fine phase mixtures as minimizers of energy.}}
\newblock \bibinfo{journal}{Arch. Ration. Mech. An.} \bibinfo{volume}{100},
  \bibinfo{pages}{13--52}.
\newblock \DOIprefix\doi{10.1007/BF00281246}.
\bibitem[{Ba\v{z}ant and Jir\'{a}sek(2002)}]{BAZ-JIR-02}
\bibinfo{author}{Ba\v{z}ant, Z.}, \bibinfo{author}{Jir\'{a}sek, M.},
  \bibinfo{year}{2002}.
\newblock \bibinfo{title}{Nonlocal integral formulations of plasticity and
  damage: Survey of progress.}
\newblock \bibinfo{journal}{J. Eng. Mech.} \bibinfo{volume}{128},
  \bibinfo{pages}{1119--1149}.
\bibitem[{Bechle and Kyriakides(2014)}]{BEC-KYR-14}
\bibinfo{author}{Bechle, N.}, \bibinfo{author}{Kyriakides, S.},
  \bibinfo{year}{2014}.
\newblock \bibinfo{title}{Localization in {N}i{T}i tubes under bending}.
\newblock \bibinfo{journal}{Int. J. Solids Struct.} \bibinfo{volume}{51},
  \bibinfo{pages}{967--980}.
\bibitem[{Bechle and Kyriakides(2016)}]{BEC-KYR-16}
\bibinfo{author}{Bechle, N.}, \bibinfo{author}{Kyriakides, S.},
  \bibinfo{year}{2016}.
\newblock \bibinfo{title}{Evolution of phase transformation fronts and
  associated thermal effects in a {N}i{T}i tube under a biaxial stress state}.
\newblock \bibinfo{journal}{Extrem Mech. Letters} \bibinfo{volume}{8},
  \bibinfo{pages}{55--63}.
\bibitem[{Bian et~al.(2018)Bian, Saleha, Pereloma, Davies and
  Gazder}]{BIA-PER-18}
\bibinfo{author}{Bian, X.}, \bibinfo{author}{Saleha, A.},
  \bibinfo{author}{Pereloma, E.}, \bibinfo{author}{Davies, C.},
  \bibinfo{author}{Gazder, A.}, \bibinfo{year}{2018}.
\newblock \bibinfo{title}{A digital image correlation study of a {N}i{T}i alloy
  subjected to monotonic uniaxial and cyclic loading-unloading in tension}.
\newblock \bibinfo{journal}{Materials Sci. Eng. A} \bibinfo{volume}{726},
  \bibinfo{pages}{102--112}.
\bibitem[{Bobinski and Tejchman(2004)}]{BOB-TEJ-04}
\bibinfo{author}{Bobinski, J.}, \bibinfo{author}{Tejchman, J.},
  \bibinfo{year}{2004}.
\newblock \bibinfo{title}{Numerical simulations of localization of deformation
  in quasi-brittle materials within non-local softening plasticity}.
\newblock \bibinfo{journal}{Comp. Concrete} \bibinfo{volume}{1},
  \bibinfo{pages}{433--455}.
\bibitem[{Bobinski and Tejchman(2005)}]{BOB-TEJ-05}
\bibinfo{author}{Bobinski, J.}, \bibinfo{author}{Tejchman, J.},
  \bibinfo{year}{2005}.
\newblock \bibinfo{title}{Modelling of concrete behaviour with a non-local
  continuum damage approach}.
\newblock \bibinfo{journal}{Arch. Hydro-Eng. Environ. Mech.}
  \bibinfo{volume}{52}, \bibinfo{pages}{243--263}.
\bibitem[{Chatziathanasiou et~al.(2016)Chatziathanasiou, Chemisky,
  Chatzigeorgiou and Meragni}]{CHA-CHE-16}
\bibinfo{author}{Chatziathanasiou, D.}, \bibinfo{author}{Chemisky, Y.},
  \bibinfo{author}{Chatzigeorgiou, G.}, \bibinfo{author}{Meragni, F.},
  \bibinfo{year}{2016}.
\newblock \bibinfo{title}{Modeling of coupled phase transformation and
  reorientation in shape memory alloys under non-proportional thermomechanical
  loading}.
\newblock \bibinfo{journal}{Int. J. Plast.} \bibinfo{volume}{82},
  \bibinfo{pages}{192--224}.
\bibitem[{Chemisky et~al.(2011)Chemisky, Duval, Patoor and Ben~Zineb}]{CHE-DUV}
\bibinfo{author}{Chemisky, Y.}, \bibinfo{author}{Duval, A.},
  \bibinfo{author}{Patoor, E.}, \bibinfo{author}{Ben~Zineb, T.},
  \bibinfo{year}{2011}.
\newblock \bibinfo{title}{Constitutive model for shape memory alloys including
  phase transformation, martensitic reorientation and twins accommodation}.
\newblock \bibinfo{journal}{Mech. Mater.} \bibinfo{volume}{43},
  \bibinfo{pages}{361--376}.
\bibitem[{Cisse et~al.(2016a)Cisse, Zaki and Ben~Zineb}]{ZAK-BZ-16}
\bibinfo{author}{Cisse, C.}, \bibinfo{author}{Zaki, W.},
  \bibinfo{author}{Ben~Zineb, T.}, \bibinfo{year}{2016}a.
\newblock \bibinfo{title}{A review of constitutive models and modeling
  techniques for shape memory alloys}.
\newblock \bibinfo{journal}{Int. J. Plast.} \bibinfo{volume}{76},
  \bibinfo{pages}{244--284}.
\bibitem[{Cisse et~al.(2016b)Cisse, Zaki and Ben~Zineb}]{ZAK-BZ-16b}
\bibinfo{author}{Cisse, C.}, \bibinfo{author}{Zaki, W.},
  \bibinfo{author}{Ben~Zineb, T.}, \bibinfo{year}{2016}b.
\newblock \bibinfo{title}{A review of modeling techniques for advanced effects
  in shape memory alloy behavior}.
\newblock \bibinfo{journal}{Smart Mater. Struct.} \bibinfo{volume}{25},
  \bibinfo{pages}{103001}.
\bibitem[{Dacorogna(2008)}]{dacorogna}
\bibinfo{author}{Dacorogna, B.}, \bibinfo{year}{2008}.
\newblock \bibinfo{title}{Direct Methods in the Calculus of Variations}.
  volume~\bibinfo{volume}{78} of \textit{\bibinfo{series}{Applied Mathematical
  Sciences}}.
\newblock \bibinfo{edition}{2nd} ed., \bibinfo{publisher}{Springer-Verlag, New
  York}.
\bibitem[{Duval et~al.(2011)Duval, Haboussi and Ben~Zineb}]{DUV-CHE}
\bibinfo{author}{Duval, A.}, \bibinfo{author}{Haboussi, M.},
  \bibinfo{author}{Ben~Zineb, T.}, \bibinfo{year}{2011}.
\newblock \bibinfo{title}{Modelling of localization and propagation of phase
  transformation in superelastic {SMA} by a gradient nonlocal approach}.
\newblock \bibinfo{journal}{Int. J. Solids Struct.} \bibinfo{volume}{48},
  \bibinfo{pages}{1879--1893}.
\bibitem[{Elibol and Wagner(2015)}]{ELI-WAG}
\bibinfo{author}{Elibol, C.}, \bibinfo{author}{Wagner, M.X.},
  \bibinfo{year}{2015}.
\newblock \bibinfo{title}{Investigation of the stress-induced martensitic
  transformation in pseudoelastic {N}i{T}i under uniaxial tension, compression
  and compression-shear}.
\newblock \bibinfo{journal}{Mater. Sci. Eng. A} \bibinfo{volume}{621},
  \bibinfo{pages}{76--81}.
\bibitem[{Eshelby(1957)}]{ESH-57}
\bibinfo{author}{Eshelby, J.}, \bibinfo{year}{1957}.
\newblock \bibinfo{title}{The determination of the elastic field of an
  ellipsoidal inclusion, and related problems}.
\newblock \bibinfo{journal}{Proc. R. Soc. London, Ser. A}
  \bibinfo{volume}{241}, \bibinfo{pages}{376--396}.
\bibitem[{Francfort and Mielke(2006)}]{francfort-mielke}
\bibinfo{author}{Francfort, G.}, \bibinfo{author}{Mielke, A.},
  \bibinfo{year}{2006}.
\newblock \bibinfo{title}{Existence results for a class of rate-independent
  material models with nonconvex elastic energies}.
\newblock \bibinfo{journal}{J. Reine Angew. Math.} \bibinfo{volume}{595},
  \bibinfo{pages}{55--91}.
\bibitem[{Frost et~al.(2016a)Frost, Bene{\v{s}}ov{\'a} and
  Sedl{\'a}k}]{FRO-BEN-MMS}
\bibinfo{author}{Frost, M.}, \bibinfo{author}{Bene{\v{s}}ov{\'a}, B.},
  \bibinfo{author}{Sedl{\'a}k, P.}, \bibinfo{year}{2016}a.
\newblock \bibinfo{title}{A microscopically motivated constitutive model for
  shape memory alloys: formulation, analysis and computations}.
\newblock \bibinfo{journal}{Math. Mech. Solids} \bibinfo{volume}{21},
  \bibinfo{pages}{358--382}.
\bibitem[{Frost et~al.(2018a)Frost, Sedl\'{a}k and Ben~Zineb}]{APPA-LOC}
\bibinfo{author}{Frost, M.}, \bibinfo{author}{Sedl\'{a}k, P.},
  \bibinfo{author}{Ben~Zineb, T.}, \bibinfo{year}{2018}a.
\newblock \bibinfo{title}{Experimental observations and modeling of
  localization in superelastic {N}i{T}i polycrystalline alloys: State of the
  art}.
\newblock \bibinfo{journal}{Acta Phys. Pol. A} \bibinfo{volume}{134},
  \bibinfo{pages}{847--}.
\bibitem[{Frost et~al.(2018b)Frost, Sedl\'{a}k, Heller, Kade{\v{r}}{\'a}vek and
  {\v S}ittner}]{SMS-SNAKE}
\bibinfo{author}{Frost, M.}, \bibinfo{author}{Sedl\'{a}k, P.},
  \bibinfo{author}{Heller, L.}, \bibinfo{author}{Kade{\v{r}}{\'a}vek, L.},
  \bibinfo{author}{{\v S}ittner, P.}, \bibinfo{year}{2018}b.
\newblock \bibinfo{title}{Experimental and computational study on phase
  transformations in superelastic {N}i{T}i snake-like spring}.
\newblock \bibinfo{journal}{Smart Mater. Struct.} \bibinfo{volume}{27},
  \bibinfo{pages}{095005}.
\bibitem[{Frost et~al.(2016b)Frost, Sedl{\'a}k, Kade{\v{r}}{\'a}vek, Heller and
  {\v{S}}ittner}]{JIMSS-SPRING}
\bibinfo{author}{Frost, M.}, \bibinfo{author}{Sedl{\'a}k, P.},
  \bibinfo{author}{Kade{\v{r}}{\'a}vek, L.}, \bibinfo{author}{Heller, L.},
  \bibinfo{author}{{\v{S}}ittner, P.}, \bibinfo{year}{2016}b.
\newblock \bibinfo{title}{Modeling of mechanical response of {N}i{T}i shape
  memory alloy subjected to combined thermal and non-proportional mechanical
  loading: a case study on helical spring actuator}.
\newblock \bibinfo{journal}{J. Intel. Mat. Syst. Str.} \bibinfo{volume}{27},
  \bibinfo{pages}{1927--1938}.
\bibitem[{Frost et~al.(2014)Frost, Sedl\'{a}k, Kruisov\'{a} and
  Landa}]{JMEP-STENT}
\bibinfo{author}{Frost, M.}, \bibinfo{author}{Sedl\'{a}k, P.},
  \bibinfo{author}{Kruisov\'{a}, A.}, \bibinfo{author}{Landa, M.},
  \bibinfo{year}{2014}.
\newblock \bibinfo{title}{Simulations of self-expanding braided stent using
  macroscopic model of {N}i{T}i shape memory alloys covering {R}-phase}.
\newblock \bibinfo{journal}{J. Mater. Eng. Perform.} \bibinfo{volume}{23},
  \bibinfo{pages}{2584--2590}.
\bibitem[{Frost et~al.(2018c)Frost, Sedl\'{a}k, Sedm{\'a}k, Heller and {\v
  S}ittner}]{SMST-LOC}
\bibinfo{author}{Frost, M.}, \bibinfo{author}{Sedl\'{a}k, P.},
  \bibinfo{author}{Sedm{\'a}k, P.}, \bibinfo{author}{Heller, L.},
  \bibinfo{author}{{\v S}ittner, P.}, \bibinfo{year}{2018}c.
\newblock \bibinfo{title}{{SMA} constitutive modeling backed up by {3D-XRD}
  experiments: Transformation front in stretched {N}i{T}i wire}.
\newblock \bibinfo{journal}{Shap. Mem. Superelasticity} \bibinfo{volume}{4},
  \bibinfo{pages}{411--416}.
\bibitem[{Grossman et~al.(2010)Grossman, Schaefer and Wagner}]{GRO-WAG-10}
\bibinfo{author}{Grossman, C.}, \bibinfo{author}{Schaefer, A.},
  \bibinfo{author}{Wagner, M.F.X.}, \bibinfo{year}{2010}.
\newblock \bibinfo{title}{A finite element study on localized deformation and
  functional fatigue in pseudoelastic {N}i{T}i strips}.
\newblock \bibinfo{journal}{Mater. Sci. Eng. A} \bibinfo{volume}{527},
  \bibinfo{pages}{1172--1178}.
\bibitem[{Gu et~al.(2015)Gu, Zaki, Morin, Moumni and Zhang}]{GU-ZAK}
\bibinfo{author}{Gu, X.}, \bibinfo{author}{Zaki, W.}, \bibinfo{author}{Morin,
  C.}, \bibinfo{author}{Moumni, Z.}, \bibinfo{author}{Zhang, W.},
  \bibinfo{year}{2015}.
\newblock \bibinfo{title}{Time integration and assessment of a model for shape
  memory alloys considering multiaxial nonproportional loading cases}.
\newblock \bibinfo{journal}{Int. J. Solids Struct.} \bibinfo{volume}{54},
  \bibinfo{pages}{28--99}.
\bibitem[{Hallai and Kyriakides(2013)}]{HAL-KYR}
\bibinfo{author}{Hallai, J.F.}, \bibinfo{author}{Kyriakides, S.},
  \bibinfo{year}{2013}.
\newblock \bibinfo{title}{Underlying material response for l{\"u}ders-like
  instabilities}.
\newblock \bibinfo{journal}{Int. J. Plast.} \bibinfo{volume}{47},
  \bibinfo{pages}{1--12}.
\bibitem[{Halphen and Nguyen(1975)}]{HAL-NGU}
\bibinfo{author}{Halphen, B.}, \bibinfo{author}{Nguyen, Q.S.},
  \bibinfo{year}{1975}.
\newblock \bibinfo{title}{Sur les mat\'{e}riaux standard
  g\'{e}n\'{e}ralis\'{e}s}.
\newblock \bibinfo{journal}{J. Mecanique} \bibinfo{volume}{14},
  \bibinfo{pages}{39--63}.
\bibitem[{Iadicola and Shaw(2004)}]{IAD-SHA}
\bibinfo{author}{Iadicola, M.A.}, \bibinfo{author}{Shaw, J.A.},
  \bibinfo{year}{2004}.
\newblock \bibinfo{title}{Rate and thermal sensitivities of unstable
  transformation behavior in a shape memory alloy}.
\newblock \bibinfo{journal}{Int. J. Plast.} \bibinfo{volume}{20},
  \bibinfo{pages}{577--605}.
\bibitem[{Jiang et~al.(2017a)Jiang, Kyriakides, Bechle and
  Landis}]{JIA-KYR-LAN-17}
\bibinfo{author}{Jiang, D.}, \bibinfo{author}{Kyriakides, S.},
  \bibinfo{author}{Bechle, N.J.}, \bibinfo{author}{Landis, C.M.},
  \bibinfo{year}{2017}a.
\newblock \bibinfo{title}{Bending of pseudoelastic {N}i{T}i tubes}.
\newblock \bibinfo{journal}{Int. J. Solids Struct.} \bibinfo{volume}{124},
  \bibinfo{pages}{192--214}.
\bibitem[{Jiang et~al.(2017b)Jiang, Kyriakides and Landis}]{JIA-KYR-LAN-17b}
\bibinfo{author}{Jiang, D.}, \bibinfo{author}{Kyriakides, S.},
  \bibinfo{author}{Landis, C.M.}, \bibinfo{year}{2017}b.
\newblock \bibinfo{title}{Propagation of phase transformation fronts in
  pseudoelastic niti tubes under uniaxial tension}.
\newblock \bibinfo{journal}{Extrem Mech. Letters} \bibinfo{volume}{15},
  \bibinfo{pages}{113--121}.
\bibitem[{Jiang et~al.(2017c)Jiang, Kyriakides, Landis and
  Kazinakis}]{JIA-KYR-17}
\bibinfo{author}{Jiang, D.}, \bibinfo{author}{Kyriakides, S.},
  \bibinfo{author}{Landis, C.M.}, \bibinfo{author}{Kazinakis, K.},
  \bibinfo{year}{2017}c.
\newblock \bibinfo{title}{Modeling of propagation of phase transformation
  fronts in {N}i{T}i under uniaxial tension}.
\newblock \bibinfo{journal}{Eur. J. Mech. A} \bibinfo{volume}{64},
  \bibinfo{pages}{131--142}.
\bibitem[{Jir\'{a}sek and Rolshoven(2003)}]{JIR-ROL-03}
\bibinfo{author}{Jir\'{a}sek, M.}, \bibinfo{author}{Rolshoven, S.},
  \bibinfo{year}{2003}.
\newblock \bibinfo{title}{Comparison of integral-type nonlocal plasticity
  models for strain softening materials}.
\newblock \bibinfo{journal}{Int. J. Eng. Sci.} \bibinfo{volume}{41},
  \bibinfo{pages}{1553--1602}.
\bibitem[{Lagoudas et~al.(2012)Lagoudas, Hartl, Chemisky, Machado and
  Popov}]{LAG-CHE}
\bibinfo{author}{Lagoudas, D.C.}, \bibinfo{author}{Hartl, D.J.},
  \bibinfo{author}{Chemisky, Y.}, \bibinfo{author}{Machado, L.G.},
  \bibinfo{author}{Popov, P.}, \bibinfo{year}{2012}.
\newblock \bibinfo{title}{Constitutive model for the numerical analysis of
  phase transformation in polycrystalline shape memory alloys}.
\newblock \bibinfo{journal}{Int. J. Plast.} \bibinfo{volume}{32--33},
  \bibinfo{pages}{155--183}.
\bibitem[{Mao et~al.(2010)Mao, Luo, Zhang, Wub, Liu and Han}]{MAO-LIU}
\bibinfo{author}{Mao, S.C.}, \bibinfo{author}{Luo, J.F.},
  \bibinfo{author}{Zhang, Z.}, \bibinfo{author}{Wub, M.H.},
  \bibinfo{author}{Liu, Y.}, \bibinfo{author}{Han, X.D.}, \bibinfo{year}{2010}.
\newblock \bibinfo{title}{{EBSD} studies of the stress-induced {B2-B19'}
  martensitic transformation in {N}i{T}i tubes under uniaxial tension and
  compression}.
\newblock \bibinfo{journal}{Acta Mater.} \bibinfo{volume}{58},
  \bibinfo{pages}{3357--3366}.
\bibitem[{Matsumoto et~al.(1987)Matsumoto, Miyazaki and Tamura}]{MAT-OTS-87}
\bibinfo{author}{Matsumoto, O.}, \bibinfo{author}{Miyazaki, Sand~Otsuka, K.},
  \bibinfo{author}{Tamura, H.}, \bibinfo{year}{1987}.
\newblock \bibinfo{title}{Crystallography of martensitic transformation in
  {T}i--{N}i single crystals}.
\newblock \bibinfo{journal}{Acta Mettal.} \bibinfo{volume}{35},
  \bibinfo{pages}{2137--2144}.
\bibitem[{Mehrabi et~al.(2014)Mehrabi, Kadkhodaei and Elahinia}]{MEH-ELA-14b}
\bibinfo{author}{Mehrabi, R.}, \bibinfo{author}{Kadkhodaei, M.},
  \bibinfo{author}{Elahinia, M.}, \bibinfo{year}{2014}.
\newblock \bibinfo{title}{Constitutive modeling of tension-torsion coupling and
  tension-compression asymmetry in {N}i{T}i shape memory alloys}.
\newblock \bibinfo{journal}{Smart Mater. Struct.} \bibinfo{volume}{23},
  \bibinfo{pages}{075021}.
\bibitem[{Mielke and Theil(2004)}]{mielke-theil}
\bibinfo{author}{Mielke, A.}, \bibinfo{author}{Theil, F.},
  \bibinfo{year}{2004}.
\newblock \bibinfo{title}{{On rate-independent hysteresis models.}}
\newblock \bibinfo{journal}{NODEA-Nonlinear. Diff.} \bibinfo{volume}{11},
  \bibinfo{pages}{151--189}.
\newblock \DOIprefix\doi{10.1007/s00030-003-1052-7}.
\bibitem[{Mohd~Jani et~al.(2014)Mohd~Jani, Leary, Subic and Gibson}]{JAN-GIB}
\bibinfo{author}{Mohd~Jani, J.}, \bibinfo{author}{Leary, M.},
  \bibinfo{author}{Subic, A.}, \bibinfo{author}{Gibson, M.A.},
  \bibinfo{year}{2014}.
\newblock \bibinfo{title}{A review of shape memory alloy research, applications
  and opportunities}.
\newblock \bibinfo{journal}{Mater. Des.} \bibinfo{volume}{56},
  \bibinfo{pages}{1078--1113}.
\bibitem[{Mori and Tanaka(1973)}]{MOR-TAN-73}
\bibinfo{author}{Mori, T.}, \bibinfo{author}{Tanaka, K.}, \bibinfo{year}{1973}.
\newblock \bibinfo{title}{Average stress in the matrix and average elastic
  energy of materials with misfitting inclusions}.
\newblock \bibinfo{journal}{Acta Mettal.} \bibinfo{volume}{21},
  \bibinfo{pages}{571--574}.
\bibitem[{Otsuka and Wayman(1998)}]{OTS-WAY}
\bibinfo{author}{Otsuka, K.}, \bibinfo{author}{Wayman, C.M.},
  \bibinfo{year}{1998}.
\newblock \bibinfo{title}{{S}hape {M}emory {M}aterials}.
\newblock \bibinfo{publisher}{Cambridge University Press}.
\bibitem[{Peerlings et~al.(2001)Peerlings, Geers, De~Borst and
  Brekelmans}]{PEE-GEE-01}
\bibinfo{author}{Peerlings, R.}, \bibinfo{author}{Geers, M.},
  \bibinfo{author}{De~Borst, R.}, \bibinfo{author}{Brekelmans, W.},
  \bibinfo{year}{2001}.
\newblock \bibinfo{title}{A critical comparison of nonlocal and
  gradient-enhanced softening continua}.
\newblock \bibinfo{journal}{Int. J. Solids Struct.} \bibinfo{volume}{38},
  \bibinfo{pages}{7723--7746}.
\bibitem[{Peng et~al.(2008)Peng, Pi and Fan}]{PEN-FAN-08}
\bibinfo{author}{Peng, X.}, \bibinfo{author}{Pi, W.}, \bibinfo{author}{Fan,
  J.}, \bibinfo{year}{2008}.
\newblock \bibinfo{title}{A microstructure-based constitutive model for the
  pseudoelastic behavior of {N}i{T}i {SMA}s}.
\newblock \bibinfo{journal}{Int. J. Plast.} \bibinfo{volume}{24},
  \bibinfo{pages}{966--990}.
\bibitem[{Pieczyska et~al.(2013)Pieczyska, Tobushi and Kulasinski}]{PIE-TOB-13}
\bibinfo{author}{Pieczyska, E.}, \bibinfo{author}{Tobushi, H.},
  \bibinfo{author}{Kulasinski, K.}, \bibinfo{year}{2013}.
\newblock \bibinfo{title}{Development of transformation bands in {T}i{N}i {SMA}
  for various stress and strain rates studied by a fast and sensitive infrared
  camera}.
\newblock \bibinfo{journal}{Smart Mater. Struct.} \bibinfo{volume}{22},
  \bibinfo{pages}{035007}.
\bibitem[{Pouya et~al.(2017)Pouya, Elibol and Wagner}]{POU-WAG-17}
\bibinfo{author}{Pouya, M.}, \bibinfo{author}{Elibol, C.},
  \bibinfo{author}{Wagner, M.F.X.}, \bibinfo{year}{2017}.
\newblock \bibinfo{title}{Understanding complex stress states in pseudoelastic
  shape memory alloys: macroscopic modeling considering localization and
  tension-compression asymmetry}, in: \bibinfo{booktitle}{ASM International -
  International Conference on Shape Memory and Superelastic Technologies, SMST
  2017}, pp. \bibinfo{pages}{167--168}.
\bibitem[{Razaee-Hajidehi et~al.(2019)Razaee-Hajidehi, Tuma and
  Stupkiewicz}]{MRH-STU-19}
\bibinfo{author}{Razaee-Hajidehi, M.}, \bibinfo{author}{Tuma, K.},
  \bibinfo{author}{Stupkiewicz, S.}, \bibinfo{year}{2019}.
\newblock \bibinfo{title}{Gradient-enhanced thermomechanical {3D} model for
  simulation of transformation patterns in pseudoelastic shape memory alloys}.
\newblock \bibinfo{journal}{Int. J. Plast.} , \bibinfo{pages}{in press}.
\bibitem[{Reedlunn et~al.(2014)Reedlunn, Churchill, Nelson, Shaw and
  Daly}]{REE-DAL-14}
\bibinfo{author}{Reedlunn, B.}, \bibinfo{author}{Churchill, C.B.},
  \bibinfo{author}{Nelson, E.E.}, \bibinfo{author}{Shaw, J.A.},
  \bibinfo{author}{Daly, S.H.}, \bibinfo{year}{2014}.
\newblock \bibinfo{title}{Tension, compression, and bending of superelastic
  shape memory alloy tubes}.
\newblock \bibinfo{journal}{J. Mech. Phys. Solids} \bibinfo{volume}{63},
  \bibinfo{pages}{506--537}.
\bibitem[{Reedlunn et~al.(2012)Reedlunn, Daly and Shaw}]{REE-DAL-12}
\bibinfo{author}{Reedlunn, B.}, \bibinfo{author}{Daly, S.},
  \bibinfo{author}{Shaw, J.A.}, \bibinfo{year}{2012}.
\newblock \bibinfo{title}{Tension-torsion experiments on superelastic shape
  memory alloy tubes}, in: \bibinfo{booktitle}{ASME 2012 Conference on Smart
  Materials, Adaptive Structures and Intelligent Systems, SMASIS 2012}, pp.
  \bibinfo{pages}{213--222}.
\bibitem[{Reedlunn et~al.(2017)Reedlunn, Shaw and Daly}]{REE-DAL-17}
\bibinfo{author}{Reedlunn, B.}, \bibinfo{author}{Shaw, J.A.},
  \bibinfo{author}{Daly, S.H.}, \bibinfo{year}{2017}.
\newblock \bibinfo{title}{Axial-torsion behavior of superelastic {N}i{T}i
  tubes}, in: \bibinfo{booktitle}{SMST 2017: Conference Proceedings from the
  International Conference on Shape Memory and Superelastic Technologies May
  15-19, 2017, San Diego, California, USA}, \bibinfo{organization}{ASME}.
\bibitem[{Sadjadpour and Bhattacharya(2007)}]{SAD-BHA-3D}
\bibinfo{author}{Sadjadpour, A.}, \bibinfo{author}{Bhattacharya, K.},
  \bibinfo{year}{2007}.
\newblock \bibinfo{title}{A micromechanics-inspired constitutive model for
  shape-memory alloys}.
\newblock \bibinfo{journal}{Smart Mater. Struct.} \bibinfo{volume}{16},
  \bibinfo{pages}{1751--1765}.
\bibitem[{Sedl\'{a}k and Frost(2018)}]{APPA-ANISO}
\bibinfo{author}{Sedl\'{a}k, P.}, \bibinfo{author}{Frost, M.},
  \bibinfo{year}{2018}.
\newblock \bibinfo{title}{Numerical simulations of {N}i{T}i shape memory alloy
  wire behaviors in tension, compression, and torsion}.
\newblock \bibinfo{journal}{Acta Phys. Pol. A} \bibinfo{volume}{134},
  \bibinfo{pages}{842--846}.
\bibitem[{Sedl\'{a}k et~al.(2012)Sedl\'{a}k, Frost, Bene{\v s}ov{\' a}, {\v
  S}ittner and Ben~Zineb}]{SED-FRO-IJOP}
\bibinfo{author}{Sedl\'{a}k, P.}, \bibinfo{author}{Frost, M.},
  \bibinfo{author}{Bene{\v s}ov{\' a}, B.}, \bibinfo{author}{{\v S}ittner, P.},
  \bibinfo{author}{Ben~Zineb, T.}, \bibinfo{year}{2012}.
\newblock \bibinfo{title}{Thermomechanical model for {N}i{T}i-based shape
  memory alloys including {R}-phase and material anisotropy under multi-axial
  loadings}.
\newblock \bibinfo{journal}{Int. J. Plast.} \bibinfo{volume}{39},
  \bibinfo{pages}{132--151}.
\bibitem[{Sedm{\'a}k et~al.(2016)Sedm{\'a}k, Pilch, Heller, Kope{\v{c}}ek,
  Wright, Sedl{\'a}k, Frost and {\v{S}}ittner}]{SCIENCE}
\bibinfo{author}{Sedm{\'a}k, P.}, \bibinfo{author}{Pilch, J.},
  \bibinfo{author}{Heller, L.}, \bibinfo{author}{Kope{\v{c}}ek, J.},
  \bibinfo{author}{Wright, J.}, \bibinfo{author}{Sedl{\'a}k, P.},
  \bibinfo{author}{Frost, M.}, \bibinfo{author}{{\v{S}}ittner, P.},
  \bibinfo{year}{2016}.
\newblock \bibinfo{title}{Grain-resolved analysis of localized deformation in
  nickel-titanium wire under tensile load}.
\newblock \bibinfo{journal}{Science} \bibinfo{volume}{353},
  \bibinfo{pages}{559--562}.
\bibitem[{Shaw and Kyriakides(1995)}]{SHA-KYR}
\bibinfo{author}{Shaw, J.A.}, \bibinfo{author}{Kyriakides, S.},
  \bibinfo{year}{1995}.
\newblock \bibinfo{title}{Thermomechanical aspects of {N}i{T}i}.
\newblock \bibinfo{journal}{J. Mech. Phys. Solids} \bibinfo{volume}{43},
  \bibinfo{pages}{1243--1281}.
\bibitem[{Shaw and Kyriakides(1997a)}]{SHA-KYR-97b}
\bibinfo{author}{Shaw, J.A.}, \bibinfo{author}{Kyriakides, S.},
  \bibinfo{year}{1997}a.
\newblock \bibinfo{title}{Initiation and propagation of localized deformation
  in elasto-plastic strips under uniaxial tension}.
\newblock \bibinfo{journal}{Int. J. Plast.} \bibinfo{volume}{13},
  \bibinfo{pages}{837--871}.
\bibitem[{Shaw and Kyriakides(1997b)}]{SHA-KYR-97}
\bibinfo{author}{Shaw, J.A.}, \bibinfo{author}{Kyriakides, S.},
  \bibinfo{year}{1997}b.
\newblock \bibinfo{title}{On the nucleation and propagation of phase
  transformation fronts in a niti alloy}.
\newblock \bibinfo{journal}{Acta Mater.} \bibinfo{volume}{45},
  \bibinfo{pages}{683--700}.
\bibitem[{{\v S}ittner et~al.(2005){\v S}ittner, Liu and Nov{\' a}k}]{SIT-LIU}
\bibinfo{author}{{\v S}ittner, P.}, \bibinfo{author}{Liu, Y.},
  \bibinfo{author}{Nov{\' a}k, V.}, \bibinfo{year}{2005}.
\newblock \bibinfo{title}{On the origin of l\"{u}ders-like deformation of
  {N}i{T}i shape memory alloys}.
\newblock \bibinfo{journal}{J. Mech. Phys. Solids} \bibinfo{volume}{53},
  \bibinfo{pages}{1719--46}.
\bibitem[{{\v S}ittner and Nov\'{a}k(2000)}]{SIT-NOV}
\bibinfo{author}{{\v S}ittner, P.}, \bibinfo{author}{Nov\'{a}k, V.},
  \bibinfo{year}{2000}.
\newblock \bibinfo{title}{Anisotropy of martensitic transformations in modeling
  of shape memory alloy polycrystals}.
\newblock \bibinfo{journal}{Int. J. Plast.} \bibinfo{volume}{16},
  \bibinfo{pages}{1243--1268}.
\bibitem[{Stebner and Brinson(2013)}]{STE-BRI}
\bibinfo{author}{Stebner, A.P.}, \bibinfo{author}{Brinson},
  \bibinfo{year}{2013}.
\newblock \bibinfo{title}{Explicit finite element implementation of an improved
  three dimensional constitutive model for shape memory alloys}.
\newblock \bibinfo{journal}{Comput. Methods Appl. Mech. Eng.}
  \bibinfo{volume}{257}, \bibinfo{pages}{17--35}.
\bibitem[{Stupkiewicz and Petryk(2013)}]{STU-PET-12}
\bibinfo{author}{Stupkiewicz, S.}, \bibinfo{author}{Petryk, H.},
  \bibinfo{year}{2013}.
\newblock \bibinfo{title}{A robust model of pseudoelasticity in shape memory
  alloys}.
\newblock \bibinfo{journal}{Int. J. Numer. Methods Eng.} \bibinfo{volume}{93},
  \bibinfo{pages}{747--769}.
\bibitem[{Sun and Li(2002)}]{SUN-LI-02}
\bibinfo{author}{Sun, Q.P.}, \bibinfo{author}{Li, Z.Q.}, \bibinfo{year}{2002}.
\newblock \bibinfo{title}{Phase transformation in superelastic {N}i{T}i
  polycrystalline micro-tubes under tension and torsion -- from localization to
  homogeneous deformation}.
\newblock \bibinfo{journal}{Int. J. Solids Struct.} \bibinfo{volume}{39},
  \bibinfo{pages}{3797--3809}.
\bibitem[{Watkins et~al.(2018)Watkins, Reedlunn, Daly and Shaw}]{WAT-REE-18}
\bibinfo{author}{Watkins, R.}, \bibinfo{author}{Reedlunn, B.},
  \bibinfo{author}{Daly, S.H.}, \bibinfo{author}{Shaw, J.A.},
  \bibinfo{year}{2018}.
\newblock \bibinfo{title}{Uniaxial, pure bending, and column buckling
  experiments on superelastic {N}i{T}i rods and tubes}.
\newblock \bibinfo{journal}{Int. J. Solids Struct.} \bibinfo{volume}{146},
  \bibinfo{pages}{1--28}.
\bibitem[{Xiao et~al.(2017)Xiao, Zeng and Lei}]{XIA-LEI-17}
\bibinfo{author}{Xiao, Y.}, \bibinfo{author}{Zeng, P.}, \bibinfo{author}{Lei,
  L.}, \bibinfo{year}{2017}.
\newblock \bibinfo{title}{Grain size effect on mechanical performance of
  nanostructured superelastic {N}i{T}i alloy}.
\newblock \bibinfo{journal}{Mater. Res. Express} \bibinfo{volume}{4},
  \bibinfo{pages}{035702}.
\bibitem[{Young et~al.(2010)Young, Wagner, Frenzel, Schmahl and
  Eggeler}]{YOU-EGG}
\bibinfo{author}{Young, M.L.}, \bibinfo{author}{Wagner, M.F.X.},
  \bibinfo{author}{Frenzel, J.}, \bibinfo{author}{Schmahl, W.W.},
  \bibinfo{author}{Eggeler, G.}, \bibinfo{year}{2010}.
\newblock \bibinfo{title}{Phase volume fractions and strain measurements in an
  ultrafine-grained {N}i{T}i shape-memory alloy during tensile loading}.
\newblock \bibinfo{journal}{Acta Mater.} \bibinfo{volume}{58},
  \bibinfo{pages}{2344--2354}.
\bibitem[{Zhang and He(2018)}]{ZHA-HE-18}
\bibinfo{author}{Zhang, S.}, \bibinfo{author}{He, Y.}, \bibinfo{year}{2018}.
\newblock \bibinfo{title}{Fatigue resistance of branching phase-transformation
  fronts in pseudoelastic {N}i{T}i polycrystalline strips}.
\newblock \bibinfo{journal}{Int. J. Solids Struct.} \bibinfo{volume}{135},
  \bibinfo{pages}{233--244}.
\bibitem[{Zhang et~al.(2010)Zhang, Feng, He, Yu and Sun}]{ZHA-SUN-10}
\bibinfo{author}{Zhang, X.}, \bibinfo{author}{Feng, P.}, \bibinfo{author}{He,
  Y.}, \bibinfo{author}{Yu, T.}, \bibinfo{author}{Sun, Q.},
  \bibinfo{year}{2010}.
\newblock \bibinfo{title}{Experimental study on rate dependence of macroscopic
  domain and stress hysteresis in niti shape memory alloy strips}.
\newblock \bibinfo{journal}{Int. J. Mech. Sci.} \bibinfo{volume}{52},
  \bibinfo{pages}{1660--1670}.

\end{thebibliography}

\end{document}